# Cosmological Simulations with TreeSPH

Neal Katz
University of Washington, Department of Astronomy, Seattle, WA 98195

David H. Weinberg
Ohio State University, Department of Astronomy, Columbus, OH 43210 [1]

Lars Hernquist[2]
University of California, Lick Observatory, Santa Cruz, CA 95064
E-mail: nsk@astro.washington.edu, dhw@payne.mps.ohio-state.edu, lars@ucolick.edu

## ABSTRACT

We describe numerical methods for incorporating gas dynamics into cosmological simulations and present illustrative applications to the cold dark matter (CDM) scenario. Our evolution code, a version of TreeSPH (Hernquist & Katz 1989) generalized to handle comoving coordinates and periodic boundary conditions, combines smoothed–particle hydrodynamics (SPH) with the hierarchical tree method for computing gravitational forces. The Lagrangian hydrodynamics approach and individual time steps for gas particles give the algorithm a large dynamic range, which is essential for studies of galaxy formation in a cosmological context. The code incorporates radiative cooling for an optically thin, primordial composition gas in ionization equilibrium with a user-specified ultraviolet background. We adopt a phenomenological prescription for star formation that gradually turns cold, dense, Jeans-unstable gas into collisionless stars, returning supernova feedback energy to the surrounding medium. In CDM simulations, some of the baryons that fall into dark matter potential wells dissipate their acquired thermal energy and condense into clumps with roughly galactic masses. The resulting galaxy population is insensitive to assumptions about star formation; we obtain similar baryonic mass functions and galaxy correlation functions from simulations with star formation and from simulations without star formation in which we identify galaxies directly from the cold, dense gas.

---

[1]Also, Institute for Advanced Study, Princeton, NJ 08540

[2]Sloan Fellow, Presidential Faculty Fellow

## 1. Introduction

Understanding the formation and clustering of galaxies is one of the leading challenges in theoretical cosmology. It is generally believed that galaxies and large-scale structure developed through gravitational instability from small-amplitude density fluctuations present in the early universe. With some recent exceptions, most studies of structure formation in the non-linear regime have focused on the gravitational evolution of collisionless matter. However, while gravity may dominate the motion of matter on large scales, it is clear that gas dynamics, dissipation, and star formation are crucial to the formation of galaxies, which are themselves the markers by which we trace larger structures. In addition, the distribution and properties of gas on larger scales provide crucial diagnostics for probing the evolution of the intergalactic medium in clusters and voids.

This paper describes computational methods for incorporating gas dynamics and star formation into cosmological simulations. Our approach is based on TreeSPH (Hernquist & Katz 1989, hereafter HK), a Lagrangian code that combines smoothed-particle hydrodynamics (SPH, see Lucy 1977; Gingold & Monaghan 1977; Monaghan 1992) with a hierarchical tree algorithm (Barnes & Hut 1986) for computing gravitational forces. The design and tests of the basic hydro/gravity code are described in detail by HK. In this paper we discuss the changes we have made to adapt the code for cosmological simulations and the algorithms that we use to treat radiative cooling and star formation. We illustrate our methods with simulations of the cold dark matter (CDM) model (Peebles 1982; Blumenthal et al. 1984). Results from several such simulations have been reported in earlier papers (Katz, Hernquist & Weinberg 1992, hereafter KHW; Hernquist, Katz & Weinberg 1995). The methods outlined here can be applied to other gravitational instability models, with or without collisionless dark matter.

Given a cosmological model, such as CDM, simulations with gas dynamics can be used to address a variety of important questions. First, and most basic to the viability of the model, does it form galaxies and galaxy clusters? More than 15 years ago, White & Rees (1978) argued that gravitational processes alone could not produce objects resembling rich galaxy clusters because numerical simulations and analytic arguments indicated that substructure in a newly collapsing object should be erased within a few crossing times (for more recent numerical work see White et al. 1987; Gelb & Bertschinger 1994ab). White & Rees proposed that dissipation in the gas component of collapsing proto-galaxies would solve this problem, enabling the luminous parts of galaxies to reach very high densities and survive as distinct entities even long after their more loosely bound dark matter halos merge. Building on earlier work by Binney (1977), Rees & Ostriker (1977), and Silk (1977), they suggested that radiative cooling can explain the gap between the masses of the largest individual galaxies and the higher characteristic mass scale of galaxy groups and clusters. During the past few years, simulations with gas dynamics have begun to provide direct support for some of these claims by illustrating the formation of small "galaxy" clusters, in which multiple lumps of dense, neutral gas orbit within a more diffuse, common halo of hot gas and dark matter (KHW; Katz & White 1993; Evrard, Summers & Davis 1994, hereafter ESD). These results are encouraging, but many details have yet to be explored, such as the dependence of the derived galaxy luminosity function on the underlying cosmological model, on physical assumptions about cooling and star formation, and on the numerical resolution of the simulations themselves.

With a simulation that incorporates galaxy formation, one can address a second important question: how does the distribution of galaxies relate to the underlying distribution of mass? Comparisons between theoretical predictions and observed galaxy clustering depend crucially on the answer to this question, and most such comparisons either assume that galaxies trace mass or employ an *ad hoc* model of "biased" galaxy formation. [In high-resolution N-body studies one can look for features that might represent individual galaxy halos, but the transient nature of substructure in collisionless simulations makes this approach suspect, to the point where it can severely undercount galaxy halos in the high-density regions that should be the sites of the strongest galaxy clustering (Gelb & Bertschinger 1994b).] With a plausible treatment of gas dynamics and star formation, the relation between galaxies and mass becomes a prediction of a theoretical model rather than an input. Recent studies suggest that a modest "bias" does arise in the standard CDM model, with rms galaxy fluctuations about 50% stronger than rms mass fluctuations on large scales (Cen & Ostriker 1992, 1993; KHW). Further simulations with larger dynamic range are needed to demonstrate the robustness of this bias, to infer its nature and its physical cause, and to understand what happens in other scenarios for structure formation.

Hydrodynamic simulations can also address a host of questions about the fate of the gas itself. How much gas cools and is converted into stars? What are the densities and temperatures of intergalactic gas in galaxy groups, in superclusters, and in the voids that separate them? Is some form of baryonic dark matter required to reconcile the baryon density inferred from nucleosynthesis with the low observed density of luminous material?

Cosmological simulations with gas dynamics are time-consuming and intricate, but by addressing the questions above, they should ultimately permit broader and more definitive contact between theories of structure formation and observations of the local and high-redshift universe. Identifying galaxies directly in simulations removes the main source of uncertainty in comparisons with observed large-scale structure. Models that follow the process of galaxy formation and the evolution of the intergalactic gas yield predictions for the statistical distributions of galaxy properties, for measures of galaxy evolution, for absorption in quasar spectra, for the nature of voids, and for X-ray emission and microwave background distortions from galaxies, groups, clusters, and superclusters.

Relative to many problems in gas dynamics, cosmological investigations demand an enormous dynamic range in space and time. To study galaxy clustering, one would like to resolve the $\sim 10$ kpc scales of individual galaxies in boxes that are tens or hundreds of Mpc on a side. Systems with crossing times and Courant times of less than a million years must be followed for $10^{10}$ years.

The great strength of TreeSPH for cosmological applications lies in its dynamic range, which derives from its use of a Lagrangian hydrodynamics algorithm and individual particle time steps. In SPH, fluid elements are represented by particles, which respond to gravitational, pressure and viscous forces. Gas properties, including pressure gradients, are computed by averaging or "smoothing" over a fixed number of neighboring particles, typically 30–100. When the distribution of matter is homogeneous, all particles have similar smoothing volumes. However, interparticle separations shrink in collapsing regions, so smoothing lengths decrease wherever dense objects form, automatically increasing the spatial resolution in precisely those regions where a large dynamic range is needed. To take full advantage of its spatial adaptivity, TreeSPH allows particles to have individual time steps according to their physical state, so that the pace of the overall computation is not driven by the small fraction of particles that requires small time steps.

In addition to a gas component, TreeSPH can incorporate collisionless dark matter particles, which respond only to gravitational forces. Because fluid elements are also represented by particles, the gravitational treatment of the two components is identical. The time required for force computation via the hierarchical tree algorithm scales with particle number as $N \log N$, and the tree structure itself provides a natural way of finding neighbors for SPH. The efficiency of force computations done with tree algorithms does not suffer in the presence of strong non-linear clustering (*e.g.* Bouchet & Hernquist 1988), an important consideration for simulations that include enormously overdense objects like galaxies. Tree codes are gridless, so they do not impose any preferred geometry, and they can be adapted to handle a variety of boundary conditions (*e.g.* Hernquist, Bouchet & Suto 1991). The flexibility and dynamic range of TreeSPH have made it useful for a variety of astrophysical problems, ranging from collisions of binary stars (Goodman & Hernquist 1991) to galaxy interactions (Hernquist 1989; Hernquist & Mihos 1995; Barnes & Hernquist 1995; Mihos & Hernquist 1994a,b) to the collapse of proto-galaxies (Katz & Gunn 1991; Katz 1992). For the applications described in this paper, we need a cosmological version of the code that employs periodic boundaries and comoving coordinates. We discuss the details of this cosmological version in §2.

From the point of view of galaxy formation, the most important element of our simulations is radiative cooling in the gas component. Cooling enables the gas to dissipate energy within a dark matter halo, so that it can sink to the center to form dense, cold lumps, which survive as distinct entities even after their parent halos merge. Within TreeSPH, we compute cooling rates as a function of density and temperature assuming the gas to be optically thin and in ionization equilibrium with a specified ultraviolet radiation background. We discuss the validity of these assumptions and the details of our implementation in §3.

The luminous parts of galaxies consist mostly of stars, so a complete description of galaxy formation must incorporate star formation. Young stars, especially those that explode as supernovae, release copious amounts of energy, which can heat the surrounding gas. The detailed physics of star formation is poorly understood, and, in any case, our simulations have at best $\sim 10$ kpc spatial resolution, much larger than the sizes of individual star-forming regions. We must therefore resort to a simplified approach that characterizes average star formation rates on galactic scales. We discuss this procedure in §4.

In §5 we present some illustrative results from simulations of the CDM model, along the lines of those shown by KHW. Some concluding remarks follow in §6.

Cosmological hydrodynamics has become a rather active field in the past few years, with investigations by a number of different groups. At various points in this paper, we will compare our approach to those of two other groups whose interests are similar to our own. We refer to these groups as ESD (Evrard et al. 1994) and CO (Cen & Ostriker 1993) based on these representative papers, but members of both groups have worked on similar problems with other collaborators and in other combinations. ESD use an SPH code that is broadly similar to our own, but they use a particle-particle-particle-mesh ($P^3M$) algorithm instead of a tree algorithm to compute gravitational forces. CO, on the other hand, use a grid-based, Eulerian code for hydrodynamics computations, combined with a particle-mesh scheme for gravitational evolution of the dark matter component. The equivalent "methods" papers for these two codes are Evrard (1988) and Cen (1992). Kang et al. (1994) have presented a detailed comparison of simulations with identical initial conditions – and no cooling – performed by these codes, by TreeSPH, and by other Eulerian codes.

## 2. Evolution Code

### 2.1. Cosmological Variables

As detailed in this paper, our implementation differs from that of HK in many respects. Following previous workers in this field (*e.g.* Efstathiou *et al.* 1985; Evrard 1988), we integrate the equations of motion in a set of coordinates comoving with the expansion. Our formulation is completely general as regards the various cosmological parameters $\Omega$, $H_0$, and $\Lambda$ within the context of a Friedman solution to the Robertson-Walker metric. In particular, the expansion factor, $a$, evolves according to

$$\left(\frac{\dot{a}}{a}\right)^2 - \frac{8\pi}{3} G\rho = \frac{\Lambda}{3} - \frac{kc^2}{a}, \tag{1}$$

where $\rho$ is the mean mass density, $\Lambda$ is the cosmological constant, and $k = -1, 0$, or 1 for closed, flat, and open universes, respectively (*e.g.* Peebles 1980, 1993).

It is customary to express the mean mass density of the universe in terms of the critical density for $k = 0$ and $\Lambda = 0$ by $\Omega \equiv \rho/\rho_c$, where

$$\rho_c = \frac{3H^2}{8\pi G}, \tag{2}$$

and $H(t) \equiv \dot{a}/a$ is the Hubble parameter. Employing the convention that at the present time, $t_0$, the expansion factor is $a(t_0) = 1$, the solution to equation (1) for a flat universe with no cosmological constant is

$$a(t) = \left(\frac{3H_0 t}{2}\right)^{2/3}, \tag{3}$$

where $H_0 \equiv H(t_0)$ is the present value of the Hubble constant. Analogous solutions can be found for an open and closed universe with or without a cosmological constant.

With these normalizations, the redshift, $z$, is related to the expansion factor by

$$z = \frac{1}{a} - 1, \tag{4}$$

and for any of the above models with $\Lambda = 0$

$$H(t) = H_0\,(1+z)\,\sqrt{1 + \Omega_0 z}, \tag{5}$$

and

$$\Omega(t) = \Omega_0\,(1+z)/(1+\Omega_0 z), \tag{6}$$

where $\Omega_0 = \Omega(t_0)$ (Peebles 1980).

At each step in a calculation, given values for the constants $\Lambda$, $\Omega_0$, and $H_0$, we solve equation (1) for $a(t)$, and then equations (4)–(6), or their counterparts for non-zero $\Lambda$, to obtain $z$, $H(t)$, and $\Omega(t)$.

### 2.2. Collisionless Dynamics

Our simulations follow the combined evolution of gas, dark matter, and newly-formed stars in a periodic cube. It is expedient to work with coordinates comoving with the expansion, $\mathbf{x}$, which are related to proper coordinates, $\mathbf{r}$, and the expansion factor through

$$\mathbf{x} = \frac{\mathbf{r}}{a}. \tag{7}$$

The momentum of a particle with mass $m$ in these coordinates is

$$\mathbf{p} = m a^2 \dot{\mathbf{x}}, \tag{8}$$

where $\dot{\mathbf{x}} \equiv d\mathbf{x}/dt$.

We assume that the dark matter and newly-formed stars can be treated in the collisionless limit and hence evolve according to the collisionless Boltzmann equation (CBE)

$$\frac{\partial f}{\partial t} + \dot{\mathbf{x}} \cdot \nabla_{\mathbf{x}} f + a m \mathbf{g} \cdot \nabla_{\mathbf{p}} f = 0, \tag{9}$$

where $\mathbf{g}$ is the peculiar acceleration and $f(\mathbf{x}, \mathbf{p}, t)$ is the distribution function (Peebles 1980). It is not feasible to solve equation (9) using finite-differences; instead we adopt the Monte Carlo "N-body" approach in which representative elements of phase fluid, or "particles," are integrated along the characteristic curves of the CBE (for a clear discussion of the relation between the N-body method and the CBE see Leeuwin, Combes & Binney [1993]). The equations of motion for the $i$-th particle are thus

$$\dot{\mathbf{x}}_i = \frac{d\mathbf{x}_i}{dt} \tag{10}$$

and

$$\ddot{\mathbf{x}}_i + 2H\dot{\mathbf{x}}_i - \frac{1}{a}\mathbf{g}_i = 0. \tag{11}$$

Unlike Efstathiou *et al.* (1985) and Bouchet & Hernquist (1988) we employ proper cosmic time, $t$, as the independent time variable in our simulations in order to facilitate the incorporation of microphysical processes in the gas phase. These equations of motion are integrated using a time-centered leapfrog algorithm (Press *et al.* 1986). Thus, if the spatial coordinates are known at time step $n - 1/2$ and the velocities are known at time step $n$, the values of $\mathbf{x}_i$ at time step $n + 1/2$, $\mathbf{x}_i^{n+1/2}$ are given by

$$\mathbf{x}_i^{n+1/2} = \mathbf{x}_i^{n-1/2} + \delta t \ddot{\mathbf{x}}_i^n. \tag{12}$$

These new spatial coordinates are then used to estimate the peculiar acceleration at time step $n + 1/2$, making it possible to update velocities according to

$$\ddot{\mathbf{x}}_i^{n+1} = \ddot{\mathbf{x}}_i^n \frac{1 - H^{n+1/2}\delta t}{1 + H^{n+1/2}\delta t} + \frac{\delta t}{a^{n+1/2}} \frac{\mathbf{g}_i^{n+1/2}}{1 + H^{n+1/2}\delta t}, \tag{13}$$

where $a^{n+1/2}$ and $H^{n+1/2}$ are the values of the expansion factor and the Hubble parameter at the intermediate time, respectively.

The particles in our simulations move in a periodic box of fixed comoving size. Gravitational forces are computed using a hierarchical tree code (Barnes & Hut 1986), optimized for vector architectures (Hernquist 1987, 1988, 1990). Unlike the tree algorithm employed by HK, the one used for cosmological simulations has been modified to handle fully periodic boundaries (Hernquist *et al.* 1991). In this approach, the force on each particle is composed of the contribution from the other particles within the simulation box and another term which represents the force from periodic replicas of this volume. The former is computed with a conventional tree code, while the latter is supplied by a lookup table, generated at the beginning of each simulation through Ewald decomposition (Ewald 1921; Rybicki 1986; Hernquist *et al.* 1991). In the examples here, the internal forces are computed with expansions to monopole order and a value of the tolerance parameter in the range $\theta \approx 0.5 - 0.7$.

### 2.3. Gas Dynamics

The baryonic matter in the simulations is modeled using smoothed–particle hydrodynamics (SPH; Lucy 1977; Gingold & Monaghan 1977). By analogy with the N-body approach, which represents a collisionless phase fluid with discrete particles, SPH employs particles to describe the evolution of the fluid elements that comprise a gas. In addition to gravitational forces, these gas particles experience local forces from pressure gradients and from viscosity in shocks. Our implementation of SPH is similar to that of HK. In particular, the gas particles obey the same equations of motion as the collisionless particles, equations (10)–(11), but the peculiar acceleration for the gas includes a local contribution in addition to the long-range gravitational acceleration.

Densities and other local quantities in the fluid are computed by kernel estimation in the usual manner (*e.g.* Monaghan 1992). We employ the spherically symmetric spline kernel proposed by Monaghan & Lattanzio (1985), as defined by

$$W(r,h) \;=\; \frac{1}{\pi h^3} \begin{cases} 1 \,-\, (3/2)(r/h)^2 \,+\, (3/4)(r/h)^3, & \text{for } 0 \le r/h \le 1, \\ (1/4)[2 \,-\, (r/h)]^3, & \text{for } 1 \le r/h \le 2, \\ 0, & \text{for } r/h \ge 2. \end{cases} \tag{14}$$

Each particle has its own smoothing length, $h_i$, chosen so that a fixed number of neighboring particles, $\mathcal{N}_s$, are contained within its smoothing volume, which extends out to $2h_i$. Expressions involving any two particles $i$ and $j$ are symmetrized in $h_i$ and $h_j$ using the procedure defined by HK. For example, the smoothed density associated with particle $i$ is

$$\rho_i \;=\; \sum_j m_j \, \frac{1}{2} \left[ W(r_{ij}, h_i) \,+\, W(r_{ij}, h_j) \right], \tag{15}$$

where the sum is over the other particles in the system, $m_j$ is the mass of particle $j$, $h_i$ and $h_j$ are the smoothing lengths of particles $i$ and $j$, and $r_{ij} \equiv |\mathbf{r}_i - \mathbf{r}_j|$. Our implementation differs little in these respects from that of HK, except that the periodic nature of the simulation volume must be accounted for when smoothing lengths and neighbors are determined for particles near an edge of the box. In addition, following Evrard (1988) and ESD, we have introduced a lower limit to the smoothing length of any particle, so that particles at very high densities do not require exceptionally small time steps. In practice, we do not allow smoothing lengths to drop below some fraction, $f_h$, of the gravitational softening length, $\epsilon_{\rm grav}$. For the examples in this paper, we use the same $\epsilon_{\rm grav}$ for all gas particles, and we take $f_h = 1/4$.

We write the local contribution to the peculiar acceleration of the gas particles as

$$\mathbf{g}_{\rm gas} \;=\; -\frac{\nabla P}{\rho} \,+\, \mathbf{a}_{\rm visc}, \tag{16}$$

where $P$ and $\rho$ are the pressure and gas density in physical coordinates and $\mathbf{a}_{\rm visc}$ is an acceleration arising from artificial viscosity, which is included to capture shocks in the flow. In what follows, we employ a symmetrized version of the discrete representation of $\mathbf{g}_{\rm gas}$ to preserve momentum conservation in the fluid. Specifically, $\mathbf{g}_{\rm gas}$ for particle $i$ is given by

$$\mathbf{g}_{\rm gas,i} \;=\; -\sum_j m_j \left[ 2\,\frac{\sqrt{P_i P_j}}{\rho_i \rho_j} \,+\, \Pi_{ij} \right] \frac{1}{2} \left[ \nabla_i W(r_{ij}, h_i) \,+\, \nabla_i W(r_{ij}, h_j) \right], \tag{17}$$

where $P_i$, $\rho_i$, and $P_j$, $\rho_j$, are the pressure and density associated with particles $i$ and $j$, respectively, $\Pi_{ij}$ is the artificial viscosity, and it is understood that all these quantities are given by their values in physical coordinates. In practice, we compute densities and smoothing lengths in comoving coordinates and then convert them as needed into physical coordinates using the known value of the expansion factor. We have experimented with various forms of the artificial viscosity; here, we employ one that resembles a bulk viscosity,

$$\Pi_{ij} \;=\; \frac{q_i}{\rho_i^2} + \frac{q_j}{\rho_j^2} \tag{18}$$

with

$$q_i \;=\; \begin{cases} \alpha h_i \rho_i c_i |\nabla\cdot\mathbf{v}|_i \;+\; \beta h_i^2 \rho_i |\nabla\cdot\mathbf{v}|_i^2, & \text{for } |\nabla\cdot\mathbf{v}| < 0, \\ 0, & \text{for } |\nabla\cdot\mathbf{v}| \geq 0, \end{cases} \tag{19}$$

where $c_i$ is the sound speed of particle $i$. The velocity divergence $\nabla\cdot\mathbf{v}$ is estimated according to

$$\nabla\cdot\mathbf{v}_i \;=\; -\frac{1}{\rho_i}\sum_j m_j \mathbf{v}_{ij}\cdot\frac{1}{2}\left[\nabla_i W(r_{ij},h_i) \;+\; \nabla_i W(r_{ij},h_j)\right], \tag{20}$$

where $\mathbf{v}_{ij} \equiv \mathbf{v}_i - \mathbf{v}_j$. In our simulations we typically take $\alpha = 0.5$ and $\beta = 0.5$, values that give a reasonable description of shocks. This choice for the artificial viscosity is not unique; other examples can be found in HK and Monaghan (1992).

The equations of motion for the gas particles are supplemented by an energy equation, which determines the evolution of the thermal energy, $u$. The discrete form of this equation is

$$\frac{du_i}{dt} \;=\; \sum_j m_j\left[\frac{\sqrt{P_iP_j}}{\rho_i\rho_j} + \frac{1}{2}\Pi_{ij}\right]\mathbf{v}_{ij}\cdot\frac{1}{2}\left[\nabla_i W(r_{ij},h_i) \;+\; \nabla_i W(r_{ij},h_j)\right] \;+\; \frac{\mathcal{H}_i - \Lambda_i}{\rho_i}, \tag{21}$$

where $\mathcal{H}_i$ and $\Lambda_i$ are included to describe radiative heating and cooling of the gas. (We denote heating by $\mathcal{H}_i$ instead of the more common $\Gamma_i$ to avoid colliding with our notation for photoionization rates in §3 below.) The energy equation is integrated using a semi-implicit method in which heating and cooling terms are integrated implicitly and adiabatic terms are integrated explicitly using a time step that is half the shortest time step of any SPH particle (HK; see also, Monaghan & Varnas 1988). To ensure numerical stability, we damp the cooling rate so that a particle cannot radiate more than half its thermal energy in a thermal energy time step (Katz & Gunn 1991). We employ an ideal gas equation of state, $P = (\gamma-1)\rho u$, where $\gamma = 5/3$ for a monatomic gas.

### 2.4. Time steps

The cosmological equation (1) is integrated with a time step $(\Delta t)_{\rm sys}$ that we hereafter refer to as the system time step. The system time step also represents the largest step used to integrate the equations of motion for an individual collisionless or SPH particle. In fact, in most of the examples reported here (the exception being simulations that include star formation, as discussed below), we integrate the equations of motion for all collisionless particles using the system time step $(\Delta t)_{\rm sys}$. For a gravitational softening length of 20 kpc with the spline–kernel softening (14), we find from numerical experiments that 2000–4000 time steps are required to evolve the collisionless component in one of our CDM simulations accurately to redshift zero. For the commonly used Plummer softening law, with the same value of $\epsilon_{\rm grav}$, the number of required time steps would decrease to 1500–3000.

The physical conditions of the gas particles often demand integration time steps shorter than $(\Delta t)_{\rm sys}$. Following HK, we allow gas particles in our simulations to have individual time steps, differing by powers of two. The time step of any SPH particle is chosen to satisfy a modified form of the Courant condition that takes into account the artificial viscosity. For the viscosity defined above, this criterion gives a time step for each particle $i$

$$\Delta t_i = \mathcal{C} \frac{h_i}{h_i|\nabla \cdot \mathbf{v}_i| + c_i + 1.2(\alpha c_i + \beta h_i |\nabla \cdot \mathbf{v}_i|)}, \tag{22}$$

where $\mathcal{C}$ is the Courant number. Typically, we take $\mathcal{C} = 0.3$.

In order to prevent a small number of particles from consuming an inordinate amount of computational effort, we also impose a minimum gas time step, usually equal to 1/8 or 1/16 of $(\Delta t)_{\rm sys}$. When equation (22) yields a time step below this minimum, we satisfy the Courant condition by increasing the particle's smoothing length $h_i$ instead of reducing its time step. (The alternative of ignoring the Courant condition for some particles can lead to dangerously unstable results.) With our adopted minimum time step, the number of particles that have smoothing lengths increased in this way is very small, often zero. However, in a simulation with a large minimum time step, increasing smoothing lengths to satisfy the Courant condition can significantly degrade spatial resolution in some regions. The Courant criterion provides a concrete illustration of a more general point: in any simulation, hydrodynamic or gravitational, achieving high spatial resolution requires high time resolution.

We could improve the efficiency of our collisionless particle integrations by again assigning individual time steps, based on dynamical criteria. This change would allow us to use a larger $(\Delta t)_{\rm sys}$ and enforce smaller time steps only for those particles that require them. We have conducted a number of experiments with different criteria, and our most successful approach so far has been to use the minimum of $\eta\left(\frac{\eta\epsilon_{\rm grav}}{v}\right)$ and $\eta\left(\frac{\epsilon_{\rm grav}}{a}\right)^{1/2}$ where $v$ is the particle velocity, $a$ is the acceleration on the particle, and $\eta$ is a constant smaller than one. Unfortunately, these criteria are not time-symmetric, so the system does not follow a surrogate Hamiltonian, and secular changes in energy can occur (Saha & Tremaine 1992). Our experiments indicate that these secular energy changes are small, provided that we use the method of Quinn et al. (1995) to retain second-order accuracy of the integration *instead* of using the second-order correction terms listed in equation (2.42) of HK. We have therefore adopted this approach for some simulations, though we hope to shift eventually to criteria for collisionless particles that are fully time-symmetric.

The use of individual time steps for collisionless particles is especially important in simulations with star formation, since these create collisionless particles in regions where densities are extremely high and dynamical times are correspondingly short (see §4). If particle time steps are too large, dense clumps of "star" particles can evaporate over the course of the simulation because inadequate time resolution enhances spurious numerical scattering. In our simulations with star formation, therefore, we adopt the above time step criteria with $\eta = 0.4$, for both the star and dark matter particles. For gas particles we apply these dynamical criteria and the Courant criterion (22) and choose the smallest implied time step.

### 2.5. Initial Conditions

Our procedure for setting up initial conditions is similar to the standard techniques used for cosmological N-body simulations (e.g. Doroshkevich et al. 1980; Efstathiou et al. 1985). We generate a Gaussian random density field with the desired linear-theory power spectrum by choosing the real and imaginary parts of each Fourier component $\delta_{\mathbf{k}}$ randomly from independent Gaussian distributions with variance $P(k)/2$. We position particles on a uniform grid, then use the Zel'dovich (1970) approximation and the linear density field to compute initial displacements and velocities for these particles. We offset the grid of SPH particles from the grid of collisionless particles by 1/2 grid cell in each dimension, and we interpolate the Zel'dovich displacement field to obtain values at the offset positions.

Our typical initial conditions use $32^3$ or $64^3$ collisionless particles and an equal number of SPH particles, with all particles of the same species having equal mass. The ratio of the collisionless particle mass to the SPH particle mass is therefore $(\Omega - \Omega_b)/\Omega_b$, where $\Omega_b$ is the baryon density parameter. However, the scheme is quite flexible and can be adapted depending on the goal of the simulation. One can, for instance, reduce the mass ratio by using more dark matter particles than SPH particles. Alternatively, one can select a specific region of the simulation to model at high resolution and use a coarser grid of more massive collisionless particles outside the high-resolution region to provide the physically appropriate tidal forces. This nesting technique has been applied successfully by Katz & White (1993) and Katz et al. (1994).

The simulations shown in this paper begin at a redshift $z = 49$, at which time we set the temperature of all gas particles to $T = 10^4$K. Energy input from small-scale non-linear processes could rather easily maintain the gas at this temperature because radiative cooling is inefficient below $10^4$K. However, until fluctuations become non-linear on the smallest scales that the simulation can resolve, the gas cools adiabatically with the expansion of the universe. A high initial temperature would suppress the collapse of gas into small dark matter halos – below the baryon Jeans mass – but with our starting temperature the Jeans mass is below the smallest scale that we can resolve in any case, and in this limit the value of the initial temperature should make no difference to the final results. For an examination of this point in the context of spherical perturbations, see Thoul & Weinberg (1995).

## 3. Radiative Cooling

At very early times, gas and dark matter move together under the influence of gravity. This situation changes when orbit-crossing occurs; collisionless dark matter can interpenetrate, but the gas shocks and heats, converting the kinetic energy of bulk motion into thermal energy. The most important differences between gas and dark matter evolution occur when the gas cools and sinks into high-density cores inside dark matter halos. This process is the key to the formation of units that can survive as distinct entities after the merger of their dark matter halos.

### 3.1. Cooling and Heating Rates

In our simulations, we compute radiative cooling for a gas of primordial composition using the 2-body processes listed in table 3 of Black (1981; 3-body cooling processes are unimportant at the densities we are able to resolve). These processes are collisional excitation of neutral hydrogen ($H^0$) and singly ionized helium ($He^+$), collisional ionization of $H^0$, $He^0$, and $He^+$, standard recombination of $H^+$, $He^+$, and $He^{++}$, dielectronic recombination of $He^+$, and free-free emission (Bremsstrahlung). Table 1 lists our full set of adopted cooling rates. The formulae are taken from Black (1981), with the modifications introduced by Cen (1992) to enforce proper behavior at temperatures $T \gtrsim 10^5$K. We adopt c.g.s. units throughout this Section, so the cooling rates of Table 1 are listed in $\mathrm{erg\,s^{-1}\,cm^{-3}}$. For free-free emission, we use a Gaunt factor

$$g_{\rm ff} = 1.1 + 0.34 \exp\left[-(5.5 - \log_{10}T)^2/3.0\right], \tag{23}$$

which fits the data in table 3.3 of Spitzer (1978). In addition to the radiative cooling processes in Table 1, we include inverse Compton cooling off the microwave background at the rate (Ikeuchi & Ostriker 1986)

$$\Lambda_{\rm C} = 5.41 \times 10^{-36} n_e T (1+z)^4 \quad \mathrm{erg\,s^{-1}\,cm^{-3}}. \tag{24}$$

Each SPH particle carries information about its density and temperature. In order to compute the radiative cooling rates from Table 1, we need to know the abundances of the different ionic species of the gas. We compute these assuming that the gas is optically thin and in ionization equilibrium (but *not* thermal equilibrium) with a specified background of ultraviolet (UV) radiation. Equilibrium implies a balance between the the creation and destruction rates for each ionic species:

$$\Gamma_{\rm eH_0} n_{\rm e} n_{\rm H_0} + \Gamma_{\gamma \rm H_0} n_{\rm H_0} = \alpha_{\rm H_+} n_{\rm H_+} n_{\rm e}, \tag{25}$$

$$\Gamma_{\rm eHe_0} n_{\rm He_0} n_{\rm e} + \Gamma_{\gamma \rm He_0} n_{\rm He_0} = (\alpha_{\rm He_+} + \alpha_{\rm d}) n_{\rm He_+} n_{\rm e}, \tag{26}$$

$$\Gamma_{\rm eHe_+} n_{\rm He_+} n_{\rm e} + \Gamma_{\gamma \rm He_+} n_{\rm He_+} + (\alpha_{\rm He_+} + \alpha_{\rm d}) n_{\rm He_+} n_{\rm e} = \alpha_{\rm He_{++}} n_{\rm He_{++}} n_{\rm e} + \Gamma_{\rm eHe_0} n_{\rm He_0} n_{\rm e} + \Gamma_{\gamma \rm He_0} n_{\rm He_0}, \tag{27}$$

$$\alpha_{\rm He_{++}} n_{\rm He_{++}} n_{\rm e} = \Gamma_{\rm eHe_+} n_{\rm He_+} n_{\rm e} + \Gamma_{\gamma \rm He_+} n_{\rm He_+}. \tag{28}$$

| Process | Species | Rate[a] |
|---|---|---|
| Collisional excitation | $H^0$ | $7.50 \times 10^{-19}\ e^{-118348.0/T}(1+T_5^{1/2})^{-1} n_e n_{H_0}$ |
| | $He^+$ | $5.54 \times 10^{-17}\ T^{-0.397} e^{-473638.0/T}(1+T_5^{1/2})^{-1} n_e n_{He_+}$ |
| Collisional ionization | $H^0$ | $1.27 \times 10^{-21}\ T^{1/2} e^{-157809.1/T}(1+T_5^{1/2})^{-1} n_e n_{H_0}$ |
| | $He^0$ | $9.38 \times 10^{-22}\ T^{1/2} e^{-285335.4/T}(1+T_5^{1/2})^{-1} n_e n_{He_0}$ |
| | $He^+$ | $4.95 \times 10^{-22}\ T^{1/2} e^{-631515.0/T}(1+T_5^{1/2})^{-1} n_e n_{He_+}$ |
| Recombination | $H^+$ | $8.70 \times 10^{-27}\ T^{1/2} T_3^{-0.2}(1+T_6^{0.7})^{-1} n_e n_{H_+}$ |
| | $He^+$ | $1.55 \times 10^{-26}\ T^{0.3647} n_e n_{He_+}$ |
| | $He^{++}$ | $3.48 \times 10^{-26}\ T^{1/2} T_3^{-0.2}(1+T_6^{0.7})^{-1} n_e n_{He_{++}}$ |
| Dielectronic recombination | $He^+$ | $1.24 \times 10^{-13}\ T^{-1.5} e^{-470000.0/T}\left(1+0.3e^{-94000.0/T}\right) n_e n_{He_+}$ |
| free-free | all ions | $1.42 \times 10^{-27}\ g_{ff} T^{1/2}(n_{H_+} + n_{He_+} + 4n_{He_{++}}) n_e$ |

Table 1: Cooling Rates ($\mathrm{erg\,s^{-1}\,cm^{-3}}$)

[a] $T_n \equiv T/10^n\,\mathrm{K}$

| | | |
|---|---|---|
| $\alpha_{H_+}$ | = | $8.4 \times 10^{-11}\ T^{-1/2} T_3^{-0.2}(1+T_6^{0.7})^{-1}$ |
| $\alpha_{He_+}$ | = | $1.5 \times 10^{-10}\ T^{-0.6353}$ |
| $\alpha_d$ | = | $1.9 \times 10^{-3}\ \ T^{-1.5} e^{-470000.0/T}\left(1+0.3e^{-94000.0/T}\right)$ |
| $\alpha_{He_{++}}$ | = | $3.36 \times 10^{-10}\ T^{-1/2} T_3^{-0.2}(1+T_6^{0.7})^{-1}$ |
| $\Gamma_{eH_0}$ | = | $5.85 \times 10^{-11}\ T^{1/2} e^{-157809.1/T}(1+T_5^{1/2})^{-1}$ |
| $\Gamma_{eHe_0}$ | = | $2.38 \times 10^{-11}\ T^{1/2} e^{-285335.4/T}(1+T_5^{1/2})^{-1}$ |
| $\Gamma_{eHe_+}$ | = | $5.68 \times 10^{-12}\ T^{1/2} e^{-631515.0/T}(1+T_5^{1/2})^{-1}$ |

Table 2: Recombination and collisional ionization rates ($\mathrm{cm^3\,s^{-1}}$)

The left-hand sides of equations (25)–(28) express the rates at which $H^0$, $He^0$, $He^+$, and $He^{++}$, respectively, are destroyed by collisional ionization, photoionization, or recombination. The right-hand sides express the rates at which these species are created from other species by the same processes. Only two of the three equations (26)–(28) are independent. Our collisional ionization rates ($\Gamma_{\rm eH_0}$, etc.) and recombination rates ($\alpha_{\rm H_+}$, etc.) are taken from Black (1981) and Cen (1992); we list them in Table 2. Units are $\rm cm^3\,s^{-1}$, so multiplying by two powers of density yields the number of collisional ionizations or recombinations per second in each $\rm cm^3$. The photoionization rates are defined by

$$\Gamma_{\gamma i} \equiv \int_{\nu_i}^{\infty} \frac{4\pi J(\nu)}{h\nu}\sigma_i(\nu)d\nu \quad {\rm s}^{-1}, \tag{29}$$

where $J(\nu)$ is the intensity of the UV background at frequency $\nu$ (in $\rm erg\,s^{-1}\,cm^{-2}\,sr^{-1}\,Hz^{-1}$), and $\nu_i$ and $\sigma_i(\nu)$ are the threshold frequency and cross-section for photoionization of species $i$, which we take from Osterbrock (1989; equation 2.4 for $H^0$ and $He^+$, equation 2.31 for $He^0$).

In addition to the rate-balance equations (25)–(28) we have the number conservation equations

$$\begin{aligned} n_{\rm H_+} &= n_{\rm H} - n_{\rm H_0}, & (30)\\ n_{\rm e} &= n_{\rm H_+} + n_{\rm He_+} + 2n_{\rm He_{++}}, & (31)\\ (n_{\rm He_0} + n_{\rm He_+} + n_{\rm He_{++}})/n_{\rm H} &= y \;\equiv\; Y/(4-4Y), & (32) \end{aligned}$$

where $n_{\rm H} = \rho X/m_p$ is the number density of hydrogen nuclei and $Y$ is the helium abundance by mass. We adopt $X = 0.76$ and $Y = 0.24$.

For a specified density, temperature, and ionizing background spectrum, equations (25)–(28) and (30)–(32) provide six independent equations for the six unknown quantities $n_{\rm H_0}$, $n_{\rm H_+}$, $n_{\rm He_0}$, $n_{\rm He_+}$, $n_{\rm He_{++}}$, and $n_{\rm e}$. It is helpful to recast these equations into the system

$$\begin{aligned} n_{\rm H_0} &= n_{\rm H}\alpha_{\rm H_+}/(\alpha_{\rm H_+} + \Gamma_{\rm eH_0} + \Gamma_{\gamma \rm H_0}/n_{\rm e}), & (33)\\ n_{\rm H_+} &= n_{\rm H} - n_{\rm H_0}, & (34)\\ n_{\rm He_+} &= y n_{\rm H}/\left[1 + (\alpha_{\rm He_+} + \alpha_{\rm d})/(\Gamma_{\rm eHe_0} + \Gamma_{\gamma \rm He_0}/n_{\rm e}) + (\Gamma_{\rm eHe_+} + \Gamma_{\gamma \rm He_+}/n_{\rm e})/\alpha_{\rm He_{++}}\right], & (35)\\ n_{\rm He_0} &= n_{\rm He_+}(\alpha_{\rm He_+} + \alpha_{\rm d})/(\Gamma_{\rm eHe_0} + \Gamma_{\gamma \rm He_0}/n_{\rm e}), & (36)\\ n_{\rm He_{++}} &= n_{\rm He_+}(\Gamma_{\rm eHe_+} + \Gamma_{\gamma \rm He_+}/n_{\rm e})/\alpha_{\rm He_{++}} & (37)\\ n_{\rm e} &= n_{\rm H_+} + n_{\rm He_+} + 2n_{\rm He_{++}}. & (38) \end{aligned}$$

The collisional ionization and recombination rates depend only on temperature, and in the absence of photoionization ($\Gamma_{\gamma i} = 0$) it is trivial to solve these equations in sequence, substituting the result of one equation into the next where necessary. The density of each species scales in proportion to $n_{\rm H}$, and since we consider only two-body cooling processes, the total cooling rate at each temperature is strictly proportional to $n_{\rm H}^2$. In the presence of photoionization, it is easy to solve this system of equations by iteration, starting with an assumed value of $n_{\rm e} \approx n_{\rm H}$. The relative abundances of different species now depend on the density, which determines the rate at which recombination undoes the effect of photoionization. The cooling rate no longer scales simply with $n_{\rm H}^2$.

In addition to affecting abundances, photoionization injects energy into the gas, since the photo-electrons carry off residual energy. The heating rate from photoionization is

$$\mathcal{H} = n_{\rm H_0}\epsilon_{\rm H_0} + n_{\rm He_0}\epsilon_{\rm He_0} + n_{\rm He_+}\epsilon_{\rm He_+} \quad {\rm erg\,s^{-1}\,cm^{-3}}, \tag{39}$$

where

$$\epsilon_i = \int_{\nu_i}^{\infty} \frac{4\pi J(\nu)}{h\nu}\sigma_i(\nu)(h\nu - h\nu_i)d\nu \quad {\rm erg\,s^{-1}}. \tag{40}$$

### 3.2. Implementation

Given the density and temperature of an SPH particle, equations (33)–(38) yield the abundances of the various ionic species. These allow us to compute the cooling rates from the processes in Table 1 and the photoionization heating rate from equation (39), which are then used in the thermal energy equation (21). These abundances also yield the mean molecular weight $\mu$, which is used in the equation of state, an improvement over HK, where $\mu$ was assumed to be constant. For speed, we implement heating and cooling in TreeSPH via a lookup table, which lists $(\mathcal{H} - \Lambda)/n_{\rm H}^2$ as a function of density and temperature. We retain the $n_{\rm H}^2$ denominator even in the presence of photoionization so that the remaining dependence on density is not very strong, allowing more accurate interpolation along the density axis. Because we must solve an equation for thermal energy rather than for temperature *per se*, we space the lookup table evenly in intervals of $\log(T/\mu)$, and we use cubic spline interpolation to evaluate net cooling rates at intermediate values. We recompute the cooling lookup table every time the ionizing background changes – typically at each large system time step.

### 3.3. Illustrative Results

Figure 1 shows the cooling rate of a primordial composition gas in collisional equilibrium, i.e. the abundances of different ionic species are computed from equations (33)–(38) with the photoionization rates $\Gamma_{\gamma i}$ set to zero. Different line types in Figure 1 show the contributions from different physical processes. At temperatures $T \gtrsim 10^6$K the gas is fully ionized, and free-free emission is the primary source of cooling. At temperatures $T \sim 10^{4.3}$K and $T \sim 10^5$K, the cooling is dominated by collisional excitation of neutral hydrogen and singly ionized helium, respectively. Below $10^4$K the gas is entirely neutral, and collisions are almost never energetic enough to raise electrons to excited states, so the cooling rate is essentially zero.

Figure 2 shows cooling and heating curves in the presence of a photoionizing background, assumed to have a spectrum $J(\nu) = 10^{-22}(\nu_{\rm L}/\nu)\ {\rm erg\,s^{-1}\,cm^{-2}\,sr^{-1}\,Hz^{-1}}$, where $\nu_{\rm L}$ is the threshold frequency for the ionization of ground-state hydrogen. This spectrum is roughly consistent with the background that might be produced by observed quasars at $z \sim 2$ (e.g., Miralda-Escudé & Ostriker 1992). Figure 2a represents a gas density of $n_{\rm H} = 2.89 \times 10^{-6}\,{\rm cm}^{-3}$, the mean density at redshift $z = 2$ for $\Omega_b h^2 = 0.0125$ (where $h$ is the Hubble constant in units of 100 $\rm km\,s^{-1}\,Mpc^{-1}$). At this density, hydrogen and helium are almost completely photoionized at all temperatures. The only cooling processes are free-free emission and recombination of $H^+$ and $He^{++}$; the excitation and ionization processes that dominate Figure 1 in the range $10^4 - 10^{5.5}$K disappear completely. Below a temperature $T_{\rm eq} \approx 10^{4.8}$K, heating from photoionization outweighs radiative cooling, and the gas gains energy. Above this temperature, cooling dominates.

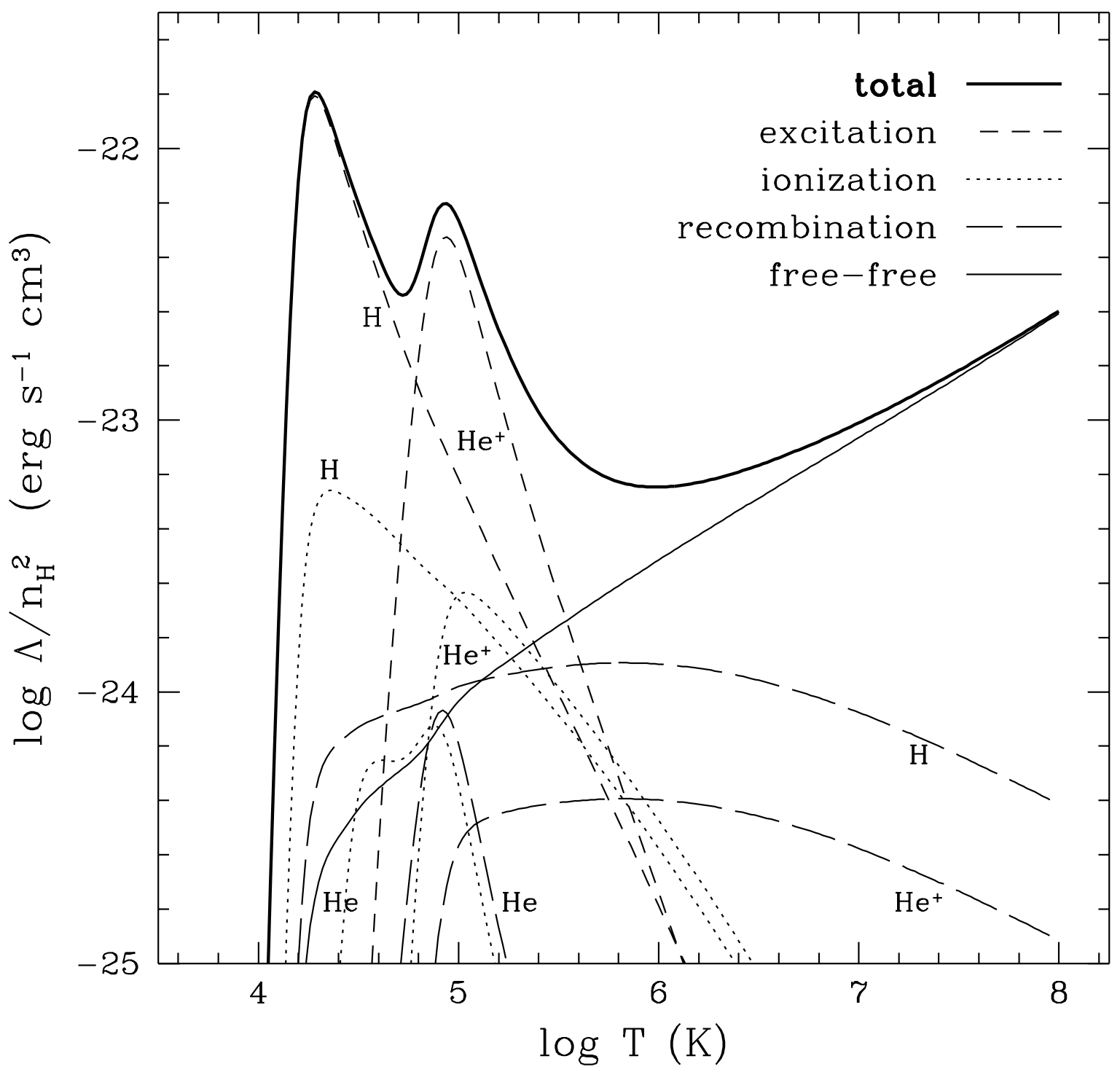

Fig. 1.— Cooling rates as a function of temperature for a primordial composition gas in collisional equilibrium. The heavy solid line shows the total cooling rate. The cooling is dominated by collisional excitation (short-dashed lines) at low temperatures and by free-free emission (thin solid line) at high temperatures. Long-dashed lines and dotted lines show the contributions of recombination and collisional ionization, respectively.

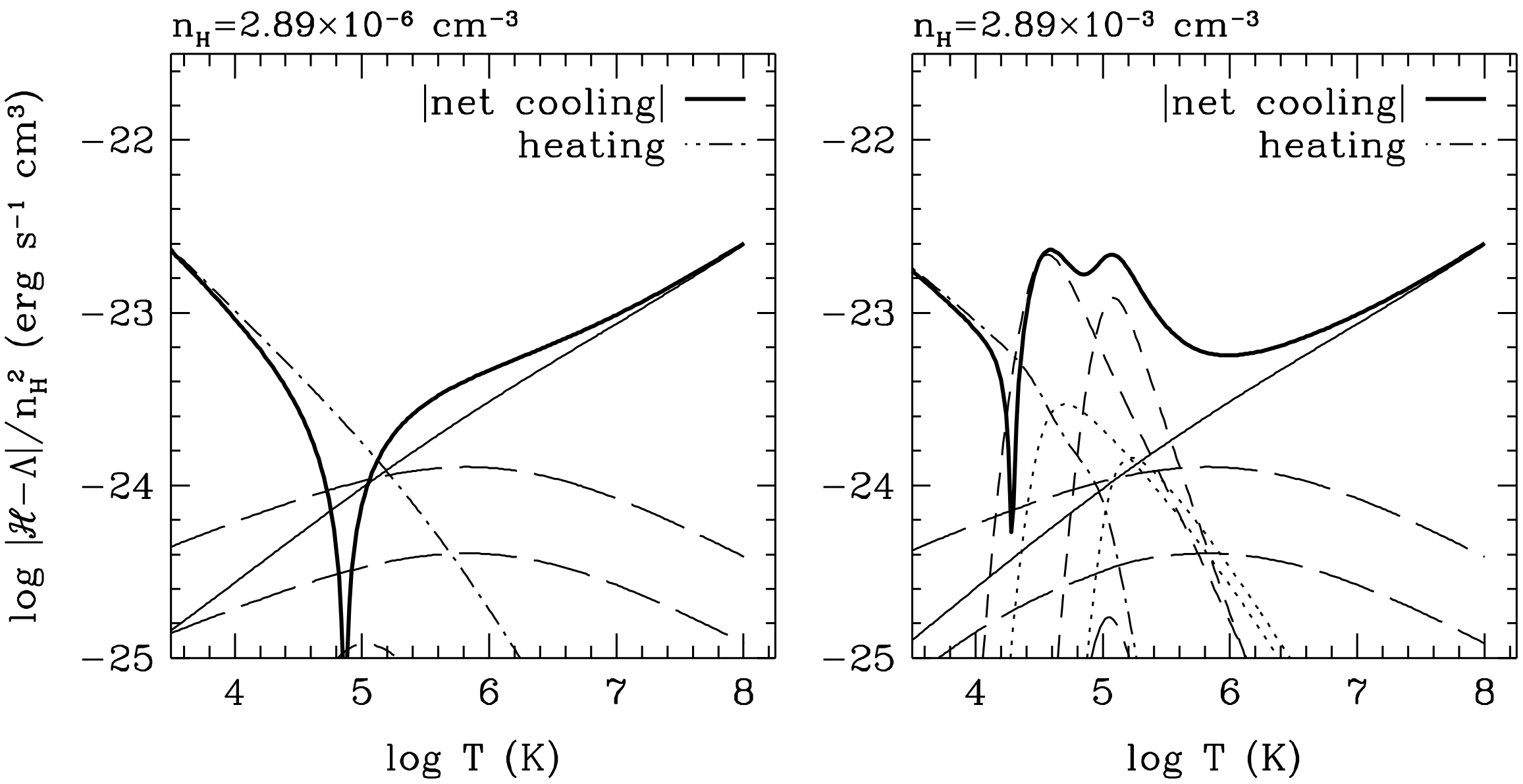

Fig. 2.— Net cooling rates as a function of temperature for primordial composition gas in ionization equilibrium with a UV radiation background of intensity $J(\nu) = 10^{-22}(\nu_{\rm L}/\nu)\ {\rm erg\,s^{-1}\,cm^{-2}\,sr^{-1}\,Hz^{-1}}$. Panel (a) corresponds to a gas density equal to the cosmic mean at $z = 2$, for $\Omega_b h^2 = 0.0125$, panel (b) to an overdensity of 1000 at $z = 2$. In each panel, the dot-dashed line shows the rate of heating by photoionization, and the heavy solid curve shows the absolute value of the net cooling rate; heating dominates at low temperatures and cooling at high temperatures. Other lines show contributions from different physical processes as in Figure 1: recombination (long-dashed), free-free (thin solid), collisional excitation (short-dashed), and collisional ionization (dotted).

Figure 2b represents a gas density 1000 times higher. At this density, recombination competes with photoionization. Abundances are closer to those of collisional equilibrium, and the cooling curve is closer to that of Figure 1. The additional cooling lowers the equilibrium temperature to $T_{\rm eq} \approx 10^{4.3}$K.

Examples of cooling curves for other combinations of UV spectrum and gas density appear in Efstathiou (1992), Thoul & Weinberg (1995), and Weinberg, Hernquist & Katz (1995). When the UV spectrum is much softer than $\nu^{-1}$ — as might be expected if the primary source of UV radiation is stars in young galaxies rather than quasars — then the helium excitation bump begins to return at densities $n_{\rm H} \sim 10^{-5}{\rm cm}^{-3}$, even though hydrogen remains fully ionized until higher densities. Equilibrium temperatures also decline somewhat for a softer spectrum because the typical photoionization event deposits less energy, though $T_{\rm eq}$ usually remains above $10^4$K. Our results for cooling rates agree well with those of Efstathiou (1992) and with independent calculations by Cen (private communication).

Figure 3 plots the distribution of SPH particles in the density-temperature plane, at redshift $z = 2$, in two simulations of the standard CDM model. The left-hand panel shows a simulation with no ionizing background. The initial conditions and parameters of this simulation are the same as those adopted by KHW ($\Omega = 1$, $H_0 = 50{\rm km\,s^{-1}\,Mpc^{-1}}$, $\Omega_b = 0.05$, $\sigma_{16,\rm mass} = 0.7$, simulation cube 22.222 comoving Mpc on a side). However, we use $64^3$ SPH particles and $64^3$ dark matter particles instead of $32^3$ particles of each species as in KHW, so the particle masses are a factor of eight smaller ($m_{\rm SPH} = 1.45 \times 10^8 M_\odot$, $m_{\rm dark} = 2.76 \times 10^9 M_\odot$).

The gas distribution in Figure 3a has three major components. One is low-density gas, which has been adiabatically cooled to low temperatures by the cosmic expansion (and, for the lowest density gas, by the expansion of voids in comoving coordinates). About 25% of the gas is below the mean density $\bar{\rho}_b$, and 30% of the gas is cooler than 100K. The second component is overdense, shock-heated gas. About 10% of the gas has $T > 10^5$K, and 5% has $T > 10^6$K. The third component consists of very overdense gas that has radiatively cooled to $T \approx 10^4$K, the location of the cutoff in the no-ionization cooling curve. About 15% of the gas is overdense by more than a factor of 1000, and 8% is overdense by more than $10^6$. It is this third component, which consists of discrete, dense lumps of cold gas, that KHW identified with galaxies. One can also discern a weak feature in the gas temperature distribution at $T \approx 10^{4.7}$K associated with the minimum in the cooling curve between the hydrogen and helium excitation bumps.

Fig. 3.— Distribution of gas in the density–temperature plane at $z = 2$ in CDM simulations with no ionizing background (a) and with an ionizing background (b). Each point represents a single SPH particle; temperatures are in degrees Kelvin and densities are scaled to the mean baryon density. Histograms show the 1-d marginal distributions, i.e. the fraction of particles in each decade of density and of temperature.

Figure 3b shows the gas distribution from a simulation with the same initial conditions but an ionizing background

$$J(\nu) = 10^{-22}(\nu_{\rm L}/\nu)F(z)\ {\rm erg\,s^{-1}\,cm^{-2}\,sr^{-1}\,Hz^{-1}}, \tag{41}$$

where the redshift-dependence is

$$F(z) = \begin{cases} 4/(1+z) & \text{for } 3 < z < 6, \\ 1 & \text{for } 2 < z < 3. \end{cases} \tag{42}$$

A drop in the ionizing background towards the highest redshifts is expected because of the decline in the number density of quasars and the increase in absorption by Lyman-alpha clouds, though our choices of a $(1+z)^{-1}$ form and an ionization redshift $z = 6$ are rather arbitrary. One can identify the same three components in this Figure as in Figure 3a, and the histogram of gas densities is nearly unchanged. However, energy input from photoionization now maintains the low-density gas at temperatures $T \sim 10^3 - 10^4$K; the more underdense gas is colder because the ratio of adiabatic cooling to photoionization heating is higher. The fraction of gas at $T \gtrsim 10^5$K is slightly higher than before, while the amount of gas in the cold, "galaxy" component is similar. This high density gas is maintained quite precisely at the equilibrium temperature where the net cooling rate is zero; this temperature varies slowly with density (see Figure 2).

### 3.4. Discussion

Photoionization can have a substantial impact on cooling rates, as a comparison of Figures 1 and 2 clearly demonstrates. Nonetheless, the photoionizing background does not change the amount of radiatively cooled gas in the CDM simulations of Figure 3, and we show elsewhere that photoionization has little effect on the masses or locations of large galaxies (Weinberg et al. 1995). Physically, this insensitivity to ionization reflects the fact that cooling times in proto-galactic gas are short even when the collisional excitation bumps of Figure 1 are absent. As a result, the collisional equilibrium approximation adopted by KHW and ESD is probably adequate for studies of large-scale galaxy clustering. Photoionization does influence the formation of low-mass galaxies because energy input heats gas before collapse, introducing a Jeans mass (see, e.g., Efstathiou 1992; Shapiro, Giroux & Babul 1994). This effect is likely to be important for systems with circular velocities $v_c \lesssim 50\,{\rm km\,s^{-1}}$(Quinn, Katz & Efstathiou 1995; Thoul & Weinberg 1995). The UV background is also essential for simulations of the Lyman-alpha forest (Cen et al. 1994; Katz et al. 1995; Hernquist et al. 1995), where photoionization determines the abundance of neutral gas.

Our treatment of radiative cooling rests on four approximations: primordial composition, ionization equilibrium, optically thin gas, and a spatially uniform radiation field. It would be straightforward to incorporate cooling from heavy elements within our framework, using the star formation algorithm described in §4 to follow the evolution of the metal abundance. However, the remaining approximations could be dropped only at a considerable cost in computational complexity and speed. The assumption of ionization equilibrium saves us from integrating stiff differential equations for several ionic species, and the other assumptions allow us to avoid laborious radiative transfer calculations.

Under the physical conditions that are relevant for most cosmological problems, the time required to reach ionization equilibrium (but *not* thermal equilibrium) is much shorter than other time scales of interest, so the abundance ratios of equations (33)-(38) should be an excellent approximation (for discussion of this point see Vedel, Hellsten & Sommer-Larsen 1994). Indeed, because the evolution equations for ionic species abundances are so stiff, it is easy to make numerical errors in direct integration that exceed the physical error of the equilibrium approximation. The assumption that gas is optically thin breaks down at high densities, but in this regime cooling rates are close to those given by collisional equilibrium. It is important to correct for self-shielding when studying the high column-density end of the Lyman-alpha forest (Katz et al. 1995), but opacity can be ignored during dynamical evolution. The assumption of a uniform radiation field is justified when the cumulative background from distant objects dominates over radiation from the closest sources. This assumption is not always valid, especially in the vicinity of bright quasars, but since photoionization has little effect on galaxy formation, it is probably safe to ignore spatial variations in the background radiation during evolution. The primary impact of fluctuations in the radiation background will be to induce variations in the Lyman-alpha forest, and to study these effects one can put in spatial variations after the fact.

Although our numerical treatment of hydrodynamics and gravity is completely different from CO's Eulerian-mesh/PM technique, our treatment of radiative cooling is quite similar. CO also adopt a uniform radiation field and optically thin gas, and while their early studies employed direct integration of abundance equations, they have recently switched to an equilibrium treatment similar to that described here (J.P. Ostriker, private communication). The major difference between our approach to radiative cooling and CO's is that we specify the history of the UV background as an input, while they compute the radiation produced by cooling gas and by galaxies and quasars formed in their simulations. CO's approach has the virtue of self-consistency: they compute the evolution of the gas in the radiation field produced by their numerical model. The disadvantage of this approach is that the heating and cooling of all the gas in the simulation becomes linked to the small-scale resolution and to necessarily speculative assumptions about quasar formation. We therefore prefer to specify the UV background externally, based on observational evidence, so that we can separate the influence of photoionization from other physical and numerical effects.

The assumptions of equilibrium, low optical depth, and spatial uniformity of the radiation field all break down during the epoch of reionization itself. In a typical model, reionization proceeds via the growth of "Stromgren spheres" around quasars and regions of active star formation, and to model this process numerically one must adopt a quite different computational framework.

Nonetheless, our assumptions should hold to reasonable accuracy after the ionization regions overlap. Miralda-Escudé & Rees (1994) point out that non-equilibrium heating during reionization can increase the entropy of the intergalactic medium, raising the post-ionization temperature somewhat above the standard equilibrium temperature. The resulting increase in the Jeans mass could have some residual impact on dwarf galaxy formation and on the lowest column-density systems in the Lyman-alpha forest.

## 4. Star Formation

### 4.1. Motivation

As shown by KHW and ESD, simulations with realistic dissipation in the baryon component form dense lumps of cold gas that can be plausibly identified with galaxies. Nonetheless, galaxies are made of stars, and it is desirable to include a description of star formation in the simulations for several reasons. Stars are collisionless, so the interactions of a stellar galaxy with other galaxies and with the intergalactic medium will differ from those of a gaseous galaxy. These differences can affect merger rates, and, as a result, they can affect the computed galaxy luminosity function. They can also influence ram-pressure stripping in groups and clusters; in particular, the destruction of a galaxy by ram pressure, seen in the simulation of Katz & White (1993), could not occur if the galaxy had already been converted to stars.

Inclusion of star formation could have important effects in the cores of clusters. In the simulation of Katz & White (1993), galaxy mergers followed by efficient dissipation lead to an extremely deep potential well at the cluster center. This deep central potential produces several effects that are contrary to observations: a rapid rise in the rotation curve towards the cluster center, a central cusp in the X–ray surface brightness profile that dominates the total X-ray luminosity, and a very large mass cooling rate. If star formation were included, the galaxies merging at the cluster center would be unable to dissipate their orbital energy through cooling radiation, so the central potential would be less deep; the more realistic physical treatment would likely produce a cluster in better agreement with observations.

Star formation returns energy to the surrounding gas through winds and radiation from supernovae and young stars. Dekel & Silk (1986) suggest that supernova feedback could slow the formation of galaxies with low circular velocities or even disrupt these galaxies entirely. White & Frenk (1991) and Cole et al. (1994) emphasize that such feedback might alter the faint end of the galaxy luminosity function, eliminating the tendency of hierarchical clustering models to produce an excess of faint galaxies. Including star formation in the simulations allows us to model the effects of feedback on the efficiency of galaxy formation and the shape of the luminosity function. Supernova feedback also limits the gas density in proto-galactic clumps, and it thus reduces the amount of angular momentum transfer that occurs as small galaxies merge to form larger ones

(Katz 1992).

A proper treatment of star formation improves the physical accuracy of the simulations, and it allows them to make contact with observations of galaxy evolution, stellar populations, and star-to-gas ratios. Our modeling of star formation is limited, unfortunately, by finite numerical resolution (typically, an individual SPH particle is more massive than a giant molecular cloud by several orders of magnitude) and by our limited understanding of the physics that governs star formation rates on galactic scales. Nonetheless, the results of KHW and ESD make the basic desideratum clear: we want an algorithm that turns dense clumps of radiatively cooled gas into stars in a smooth and physically reasonable way.

### 4.2. Star-formation criteria and star-formation rates

Our criteria for star formation and our computation of star-formation rates are similar to those of Katz (1992). Gas particles are eligible to form stars only if they are in a part of the flow that is both converging and Jeans unstable, thus ensuring that stars form only in regions that are presently collapsing and will continue to collapse. A particle is in a convergent flow if the SPH estimate for the velocity divergence is less than zero (see equation [20]). Jeans stability is determined locally by requiring that the sound crossing time be less than the gravitational dynamical time. Therefore, particle $i$ is Jeans *unstable* if

$$\frac{h_i}{c_i} > \frac{1}{\sqrt{4\pi G\rho_i}}, \qquad (43)$$

where $G$ is the gravitational constant, $c_i$ is the local sound speed, $\rho_i$ is the local gas density, and $h_i$ is the particle's SPH smoothing length. Following Summers (1994) we also require that regions of star formation have a minimum physical density corresponding to 0.1 hydrogen atoms per $\mathrm{cm}^3$. This threshold density for star formation is plausible on observational grounds (Kennicutt 1989), and it also prevents us from overestimating the effects of supernova feedback. If stars are allowed to form at lower densities, they inject energy into low-density gas that cannot cool rapidly. At high densities the cooling times are short, however, so most of the supernova energy is dissipated in cooling radiation; this situation seems to be a more realistic description of what happens in present-day galaxies, where Type II supernovae occur in high gas density environments. Our final requirement for star formation is an *over*density $\rho_g/\bar{\rho_g} > 55.7$. In an $r^{-2}$ density distribution this local density corresponds to a mean enclosed overdensity of 169, the virialization overdensity of a spherical perturbation, so this criterion restricts star formation to collapsed, virialized regions. It is important only at high redshifts, when the physical density can exceed 0.1 $\mathrm{cm}^{-3}$ in gas that is only moderately overdense.

In sum, we have four criteria that determine eligibility for star formation: a convergent flow condition, a Jeans instability requirement, a physical density threshold, and an overdensity threshold. In practice, gas that satisfies the physical density threshold usually satisfies the other conditions as well. Once a gas particle is *eligible* to form stars, its star formation *rate* is given by

$$\frac{d\rho_\star}{dt} = -\frac{d\rho_g}{dt} = \frac{c_\star \rho_g}{t_g}, \tag{44}$$

or

$$\frac{d\ln\rho_g}{dt} = -\frac{c_\star}{t_g}, \tag{45}$$

where $c_\star$ is a dimensionless star-formation rate parameter. The gas flow time scale $t_g$ is the maximum of the local gas dynamical time, $t_{\rm dyn} = (4\pi G\rho_g)^{-1/2}$, and the local cooling time; if the cooling time is shorter than the dynamical time, then the cloud can collapse unimpeded by gas pressure, but if the cooling time is longer than the local dynamical time, the cloud must wait to cool before it can fragment into stars.

We estimate the cooling time as

$$t_{\rm cool} = u_i / \frac{du_i}{dt}, \tag{46}$$

where $u_i$ is the thermal energy. In dense, star forming regions, the cooling time is typically much shorter than the dynamical time, so $t_g \propto \rho_g^{-1/2}$, and equation (44) implies a star formation rate proportional to $\rho_g^{3/2}$, similar to the observed relation for spiral galaxies (Schmidt 1959). When the gas temperature approaches $10,000$K (in the absence of photoionization) or the equilibrium temperature (with a photoionizing background), the cooling time defined by equation (46) becomes very long. Setting $t_g$ to the maximum of $t_{\rm cool}$ and $t_{\rm dyn}$ can lead to unphysical limits on the star formation rate in this regime. In fact, high-density gas at these temperatures is effectively isothermal, so thermal pressure does not impede fragmentation at all. Therefore, when the gas temperature is below $30,000$K, we always set $t_g = t_{\rm dyn}$.

### 4.3. Star particles

In the implementation of Katz (1992), gas particles "formed stars" by reducing their mass and giving birth to new, collisionless, star particles. In large-volume simulations, however, the addition of many new particles in the highest density regions can make a computation prohibitively expensive. Instead of creating new particles, we allow gas particles that form stars to temporarily maintain a dual identity as "star-gas" particles, with their mass divided between stellar and gas contributions. When computing SPH properties and forces, we use only the particles' gas masses, but the total gas+star mass contributes to gravitational forces. When the gas mass of a particle falls below 5% of its original mass, we turn the particle into a collisionless, pure-star particle. We distribute the residual gas mass and thermal energy to the surrounding gas particles in a SPH-smoothed manner. The intermediate use of dual-identity, star-gas particles allows us to incorporate star formation without either (a) artificially reducing our resolution by rapidly removing SPH particles from converging flows (as would happen if we turned gas particles instantaneously into stars), or (b) spawning large numbers of extra particles that would slow the computation and consume memory.

Equation (44) for the star formation rate implies that the probability $p$ that a gas particle forms stars in a time $\Delta t$ is

$$p = 1 - e^{-c_{\star}\Delta t/t_g}. \tag{47}$$

At each system time step, we compute $p$ from equation (47) for all eligible gas particles and draw random numbers to decide which particles actually form stars during the time step. For these particles, a fraction $\epsilon_* = 1/3$ of the gas mass is converted to stars. We also allow for a return of some of this stellar mass to interstellar gas via stellar mass loss in an instantaneous recycling approximation; recycled material is distributed amongst a particle and its neighbors in an SPH-smoothed manner. At present we assume that this recycled gas comes exclusively from supernovae, which are themselves assumed to come from stars with initial mass $M \geq 8M_{\odot}$ and to leave stellar remnants of mass $1.4M_{\odot}$. We adopt a Miller–Scalo (1979) initial mass function with a lower mass cutoff of $0.1M_{\odot}$ and an upper mass cutoff of $100M_{\odot}$. We do not presently track the metallicity of SPH gas, but it would be straightforward to add this effect or to modify our mass-loss assumptions in future work, while maintaining the basic star-formation framework outlined here.

### 4.4. Feedback

When stars form, we add supernova feedback energy to the surrounding gas in the form of heat, assuming that each supernova yields $10^{51}$ ergs. The added heat must be smoothed over the neighboring particles using the TreeSPH smoothing algorithm. Our typical system time step of $3.25 \times 10^6$ years is considerably shorter than the stellar evolution time scale of a typical supernova progenitor, so if we returned all the supernova energy in a single step we would overestimate the effects of feedback. We therefore add this energy gradually, with an exponential decay time of $2 \times 10^7$ years, the approximate lifetime of an $8M_{\odot}$ star.

Thermal energy deposited in dense, rapidly cooling gas is quickly radiated away, so although feedback has some effect in our simulations, the impact is usually not dramatic. Several authors have recently suggested that some feedback energy from supernovae should be deposited in kinetic rather than thermal form (*e.g.* Mihos & Hernquist 1994c). Navarro & White (1993), for example, find that while purely thermal feedback has little effect on surrounding gas, depositing a fraction of the feedback energy in kinetic form (by an outflow) can suppress subsequent star formation in low-mass systems.

Although we cannot be certain that adding supernova feedback energy as heat captures all of the relevant physics that goes on in a real proto-galaxy, we believe that purely thermal feedback makes the most sense in the resolution regime of current simulations. Recall that in TreeSPH (or any other SPH code), shocks are resolved over about four SPH-particle smoothing lengths, $4h$.

A supernova blast wave in a dense, uniform interstellar medium converts its kinetic energy into thermal energy through shocks. Thus, under the physical conditions implied for the star-forming gas in our simulations, we might expect an SPH simulation with infinite resolution to convert virtually all the kinetic feedback energy into thermal energy. However, with realistic numerical parameters, it is possible for kinetic energy to escape thermalization simply because the resolution of the simulation is insufficient, as the $4h$ shock resolution distance may exceed the size of the regions containing dense, star-forming gas. For our simulations, we choose SPH smoothing lengths so that each particle has $\sim 32$ neighbors within $2h$, implying an average of 256 particles within $4h$. Many of the galaxies that form in the simulation have fewer than 256 particles of cold, dense gas, so if we used a kinetic scheme we would artificially enhance the effects of feedback by not allowing it to convert properly into thermal form, from which it can be dissipated by cooling radiation.

The failure to thermalize kinetic feedback will be more severe if the minimum SPH smoothing length is limited, either indirectly, by enforcing the Courant condition while retaining a minimum time step (see §2.4), or directly, by imposing a minimum $h$ that is some fraction $f_h$ of the gravitational softening scale $\epsilon_{\rm grav}$ (see §2.3). Indeed, since clumps of dense, cold gas tend to shrink to sizes $\sim \epsilon_{\rm grav}$, a value $f_h \gtrsim 1/5$ guarantees that kinetic feedback in star-forming regions cannot be properly thermalized by shocks.

### 4.5. Examples

Figure 4 demonstrates the operation of our star formation algorithm. Each panel plots a different simulation started from the same initial conditions, which represent a CDM power spectrum in an $\Omega = 1$ universe with $H_0 = 50$ km s$^{-1}$ Mpc$^{-1}$ and $\sigma_{16} = 0.7$. The 22.222 Mpc (comoving) simulation cube is shown at redshift $z = 2$. No ionizing background is included. Panel ($a$) shows a simulation without star formation; the particles are those with density more than 1,000 times the mean baryon density and temperature $T < 30,000$K.

Panels ($b$) and ($c$) plot the star particles from simulations that include star formation, with the rate parameter $c_\star = 0.1$ and 1.0, respectively. The algorithm places stars in the high-density, cold regions that are apparent in panel ($a$). Some of these regions have not formed stars because they do not yet meet the minimum physical density criterion, but they would probably form stars at later times when their densities increase. The total mass in stars is remarkably similar in the two runs even though the $c_\star$'s differ by a factor of 10: $8.12 \times 10^{12}\, M_\odot$ for $c_\star = 0.1$ and $6.85 \times 10^{12}\, M_\odot$ for $c_\star = 1.0$. This insensitivity to the $c_\star$ parameter was also found by Katz (1992). The star formation rate increases with density, and a simulated proto-galaxy tends to contract until there is an equilibrium between self-gravity and the pressure support provided by thermal energy from supernova feedback. The value of $c_\star$ serves mainly to set this equilibrium density; the gas that cools eventually forms stars in any case. As this analysis predicts, the simulation with $c_\star = 0.1$ has denser gas than the one with $c_\star = 1.0$. The lower equilibrium gas densities in the $c_\star = 1.0$ run make cooling of the supernova energy less efficient, leading in the end to a slightly lower total stellar mass. Overall we find the insensitivity of results to our one arbitrary parameter encouraging.

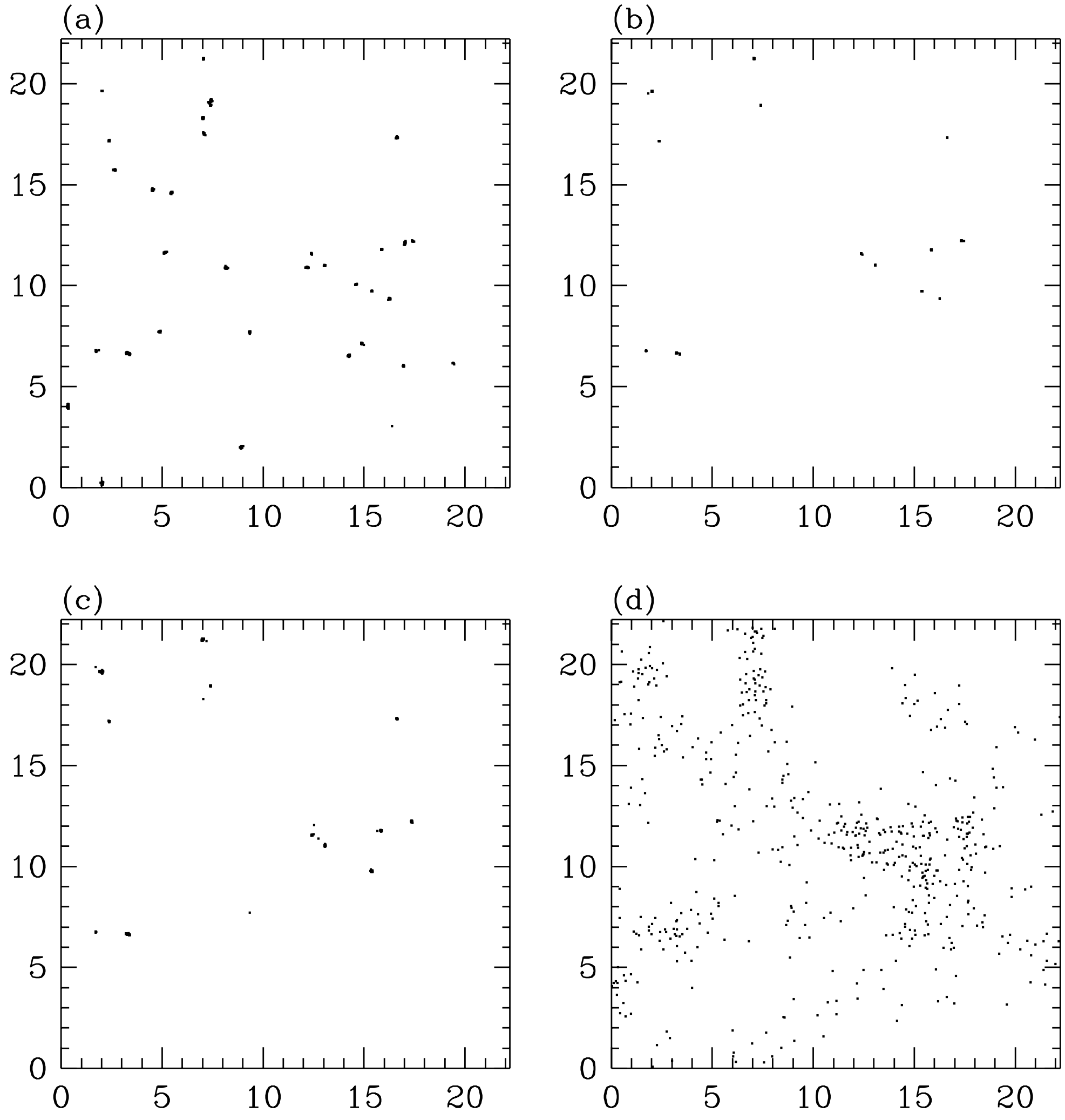

Fig. 4.— Operation of the star-formation algorithm in a CDM model. Each panel shows a projection of particles in a cube 22.222 comoving Mpc on a side, at redshift $z = 2$. (a) Gas particles with density $\rho_g > 1000\bar{\rho}_g$ and $T < 30,000$K, from a simulation without star formation. (b) Star particles from a simulation with efficiency parameter $c_* = 0.1$ and a threshold density for star formation of 0.1 atoms/cm$^3$. (c) Star particles from a simulation with $c_* = 1.0$ and a threshold of 0.1 atoms/cm$^3$. (d) Star particles from a simulation with $c_* = 0.52$ and an *over*density threshold for star formation of $\rho_g > 5.5\bar{\rho}_g$. Many of the clumps in (a)-(c) superpose large numbers of particles; panels (a)-(d) contain 1641, 960, 694, and 548 particles, respectively. Star formation in cases (b) and (c) occurs in clumps of dense, cold gas, corresponding to the more massive and denser clumps seen in (a). Because of the low density threshold for star formation in (d), the star distribution is much more diffuse.

Finally, in panel (*d*) we show a simulation where the star-formation parameters are chosen to be similar to those used by CO. We take $c_\star = 0.52$ and increase the supernova energy per unit mass of star formation by a factor of seven. We also decrease the minimum overdensity to 5.5 and remove the minimum physical density criterion. Because $\rho/\bar{\rho} > 5.5$ is a mild density criterion, stars are no longer confined to the cold, high-density regions that are apparent in the simulation without star formation. Indeed, because star formation can now occur in relatively low-density gas with long cooling times, and because the supernova energy per unit mass is higher than before, feedback suppresses the formation of dense gas knots entirely. The result is a diffuse stellar distribution without any of the concentrations of cold gas and stars that we identify as galaxies in the other simulations. The total stellar mass drops to $1.93 \times 10^{12} M_\odot$ because of the efficient feedback.

We should note that CO distribute feedback energy in time according to $f(t) \propto (t/t_{\rm dyn})\exp(-t/t_{\rm dyn})$, instead of the exponential law that we use. We expect that delaying the peak in feedback by a dynamical time would allow gas densities to increase, perhaps bringing results closer to those shown in panels (*b*) and (*c*). Unfortunately, the specifics of our implementation make it hard for us to adopt CO's time distribution instead of an exponential, so we are unable to examine this point directly. With our treatment, it seems clear that a high density and/or overdensity threshold is needed to yield realistic results.

## 5. An illustration: galaxies in the CDM model

As Figure 4 shows, hydrodynamic simulations with radiative cooling can produce dense clumps of particles that stand out from the background and are plausibly identified with galaxies. In this Section, we present results from the two simulations illustrated in Figures 4*a* and 4*b*, now evolved to redshift zero. Both simulations adopt the same CDM initial conditions as the KHW simulation. KHW evolved their simulation hydrodynamically to $z = 0.375$, then (having run out of supercomputer time) labeled cold, dense gas particles as "stars" and continued the evolution to $z = 0$ with a collisionless N-body calculation. In a sense, the two cases illustrated here bracket KHW's "instantaneous star-formation" run: the first simulation allows no conversion of gas into stars, while the second implements continuous star formation as described in §4.

Following KHW, we identify galaxies as gravitationally bound groups of particles that are associated with a common density maximum. The approach is inspired by Gelb & Bertschinger's (1994a,b) DENMAX algorithm, but instead of defining gradients of the density field on an Eulerian mesh, we use the SPH smoothing kernel to define Lagrangian density gradients. The algorithm will be described in detail elsewhere (Stadel et al. 1995). The code, together with documentation and illustrations of its operation, is available from the University of Washington HPCC group at

http://www-hpcc.astro.washington.edu.

In our gas-only simulation, we apply the galaxy identification algorithm to gas particles with temperature $T < 30,000$K and density $\rho_g/\bar{\rho}_g > 1000$. In the star-formation simulation, we apply the algorithm jointly to all star particles and to the gas particles that pass the above temperature and density thresholds. Figure 5 shows the spatial distribution of the two galaxy populations — projections of the 22.222 Mpc cube, with each galaxy represented by a circle whose area is proportional to its total baryonic mass. The two distributions are remarkably similar. There are several instances where galaxies have merged in the gas-only run while remaining distinct in the star-formation run, but in most cases one can identify galaxies in the same locations with nearly the same masses in the two simulations.

In each panel of Figure 5, the concentration of points near $x = 14.5$ Mpc, $y = 11$ Mpc, corresponds to a small cluster of galaxies in a 3-dimensional view. In both cases the cluster contains two dominant central galaxies, one with a baryonic mass of $1.5 \times 10^{12} M_\odot$ and a second of approximately half this mass. A 1.5 Mpc sphere centered on the most massive galaxy contains eight galaxies in the gas-only simulation and ten galaxies in the simulation with star formation. The total mass within this sphere is $1.3 \times 10^{14} M_\odot$, corresponding to an overdensity of $\rho/\bar{\rho} = 135$. The baryonic mass (hot gas, cold gas, and stars) is $6.3 \times 10^{12} M_\odot$; the ratio $M_b/M_{\rm tot} = 0.0485$ is just slightly below the global baryon fraction $\Omega_b = 0.05$.

Both simulations end up with 25% of the baryons in galaxies (the same result as KHW), and these galactic baryons contribute a fraction $\Omega_g = 0.012$ of closure density. In the star-formation run, 88% of this mass is stellar at $z = 0$, with the remaining 12% in cold gas. Persic & Salucci (1992) find that the mass density associated with luminous stars in observed galaxies is $\Omega_{\rm stars} \approx 0.002$, a factor of six below our simulation result. Other authors have obtained somewhat higher values, but it nonetheless seems clear that our simulation can represent the observed universe only if a large fraction of our "stars" are in fact baryonic dark matter, perhaps objects below the hydrogen-burning mass limit. Alternatively, our assumed baryon density $\Omega_b = 0.05$ may be too high, our star-formation algorithm may underestimate feedback effects, or the CDM model itself may be at fault.

Figure 6 shows the baryonic mass function of the simulated galaxy populations. The mass functions fall off below $\sim 5 \times 10^{10} M_\odot$, probably because our simulations have insufficient resolution to track the formation of smaller galaxies. The smooth solid curve represents the luminosity function of Loveday et al. (1992) — a Schechter function with parameters $\alpha = -1$, $\phi_\star = 0.00175$ Mpc$^{-3}$, and $M_\star = -21$ ($L_{B,\star} = 3.9 \times 10^{10} L_{B,\odot}$) — which we have converted to a mass function by assuming a constant ratio of baryon mass to blue light. We choose this ratio, $M_b/L_B = 10.6\ M_\odot/L_{B,\odot}$, so that the mean luminosity density produced by our simulated galaxy population matches the value implied by the luminosity function. The ratio is higher than that of a normal stellar population, another manifestation of the problem discussed in the previous paragraph. Our dynamic range in resolved galaxy masses is small, so more ambitious simulations will be needed to determine whether this model reproduces the observed form of the luminosity function.

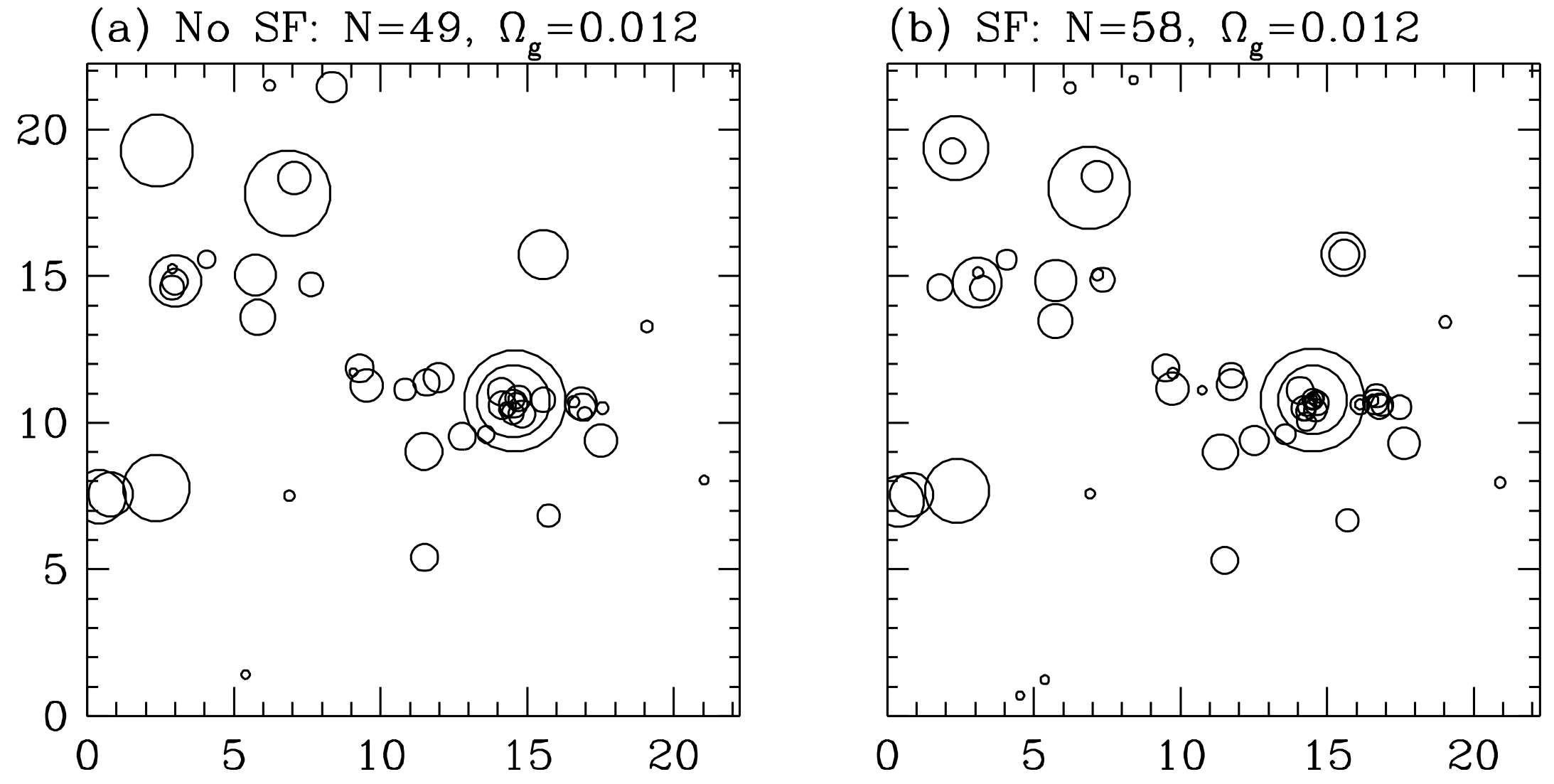

Fig. 5.— Projected distribution of "galaxies" in simulations of a CDM model, at redshift 0. (a) A simulation without star formation; galaxies are identified directly from the cold gas component. Each circle represents a galaxy, with area proportional to the galaxy's baryonic mass. (b) A simulation with star formation, in which galaxies are identified from the stars and cold gas. Each circle's area is proportional to the total (stellar+gas) baryonic mass of the corresponding galaxy. Stellar mass dominates in all but the smallest systems. The number of galaxies and the cosmic density associated with the baryonic component of galaxies is indicated above each panel.

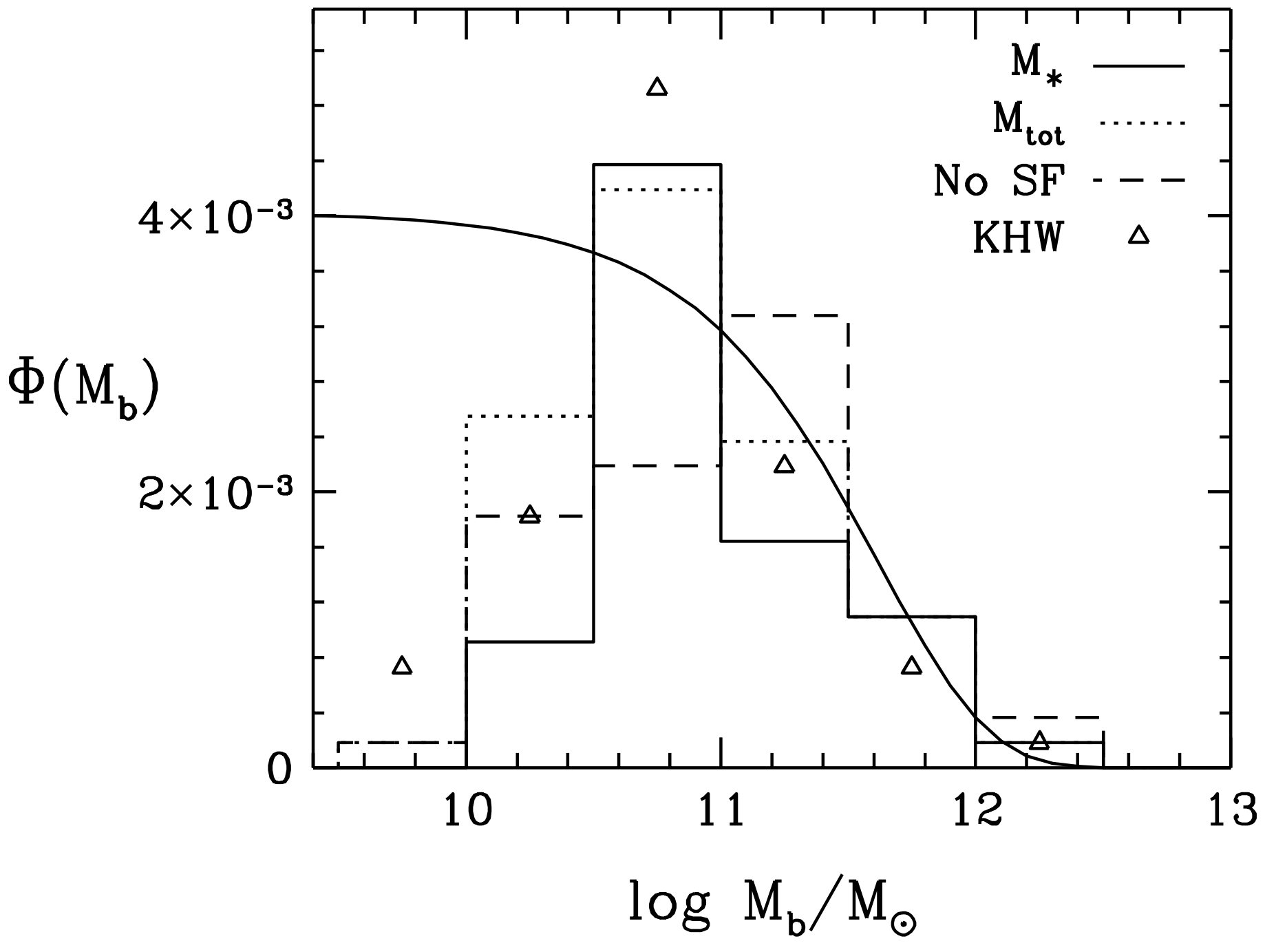

Fig. 6.— Baryonic mass function of the galaxy populations displayed in Figure 5; $\Phi(M_b)$ is the number of galaxies per $Mpc^3$ per decade of baryon mass. The solid histogram represents the distribution of galaxy stellar masses from the simulation with star formation. The dotted histogram represents the total baryonic masses, stars and cold gas. The dashed line represents the gas masses of the galaxies identified in the simulation without star formation. Triangles show the mass function obtained from the star distribution of the KHW simulation at $z = 0$. In all cases the SPH particle mass is $1.15 \times 10^9 M_\odot$, and the turnover of the mass function below $\sim 5 \times 10^{10} M_\odot$ is probably caused by resolution effects. The smooth solid curve represents the luminosity function of Loveday et al. (1992), scaled by a baryon-mass-to-blue-light ratio of $M_b/L_B = 10.6\ M_\odot/L_{B,\odot}$.

Figure 7 displays the galaxy correlation functions and the correlation function of the mass distribution from a collisionless N-body simulation with identical initial conditions. As in KHW, we find that (a) there is a weak, positive bias between the galaxy correlation function and the mass correlation function at separations of a few Mpc, and (b) the galaxy correlation function is roughly a power-law, $(r/5.3 \text{ Mpc})^{-2.1}$, in to scales $r \lesssim 100$ kpc. This power-law is steeper and of lower amplitude than the correlation function $(r/10.8 \text{ Mpc})^{-1.8}$ measured by Davis & Peebles (1983) from the CfA redshift survey, but because of our small simulation volume, we cannot readily assess the significance of this discrepancy. The absence of large-scale waves from our periodic box causes us to underestimate the amplitude of correlations at larger separations, and the small volume leads to large random errors in addition to this systematic effect. Error bars on the correlation estimates from the star-formation run show the $\sqrt{N}$ fluctuations in the number of galaxy pairs in each radial bin, but these are a lower limit to the true random errors, which also have a contribution from the finite number of independent structures in the simulation volume.

Figures 6 and 7 confirm the visual evidence of Figure 5: we obtain similar galaxy populations whether or not we convert cold gas into collisionless stars. Indeed, even KHW's much cruder treatment of star formation — instantaneous conversion of cold gas at $z = 0.375$ — yields almost identical results, as indicated by the triangular points in Figures 6 and 7. Our conclusion is at variance with that of Frenk et al. (1995; hereafter FEWS), who model the formation of a small cluster of galaxies in an $\Omega = 1$, CDM-dominated universe. They find substantial differences between a purely hydrodynamic simulation and a simulation in which cold gas is converted to collisionless stars at $z = 0.7$. They attribute these differences to strong viscous interactions near the center of the cluster in the hydrodynamic simulation. The cluster that forms in our simulations is only a factor of two less massive than the FEWS cluster, but we see no sign of such an effect.

It is possible that we would reproduce the FEWS result if we adopted their initial conditions and cosmological parameters, since the properties and formation history of their cluster are different from ours. Alternatively, the disagreement between our results may arise primarily from differences in numerical implementations. For example, FEWS use a Gaussian SPH kernel instead of the spline kernel adopted here, and the tails of the Gaussian may give galaxies a larger cross-section for viscous interactions. The FEWS simulations have better mass resolution than ours, an SPH particle mass of $3 \times 10^8 M_\odot$ vs. $1.2 \times 10^9 M_\odot$. Our simulations, on the other hand, have better time resolution. The time step for all particles in the FEWS simulations is $t_0/700$, where $t_0$ is the present age of the universe. The largest time step of particles in our simulations is $(\Delta t)_{\rm sys} = t_0/4000$ ($t_0/2000$ in the continuous star-formation run, which has individual time steps for collisionless particles), and the minimum allowed time step for SPH particles is a factor of 8–16 smaller. Without further experiments, it is difficult to know how these numerical parameters affect the mergers and disruptions of gaseous and stellar galaxies inside a cluster. Our simulations certainly lead to a happier conclusion — that the simulated galaxy population is insensitive to specific assumptions about star formation — but that does not necessarily mean that they are more physically accurate. The issue is an important one for studies of galaxy cluster formation, and it warrants further investigation with more carefully tailored simulations.

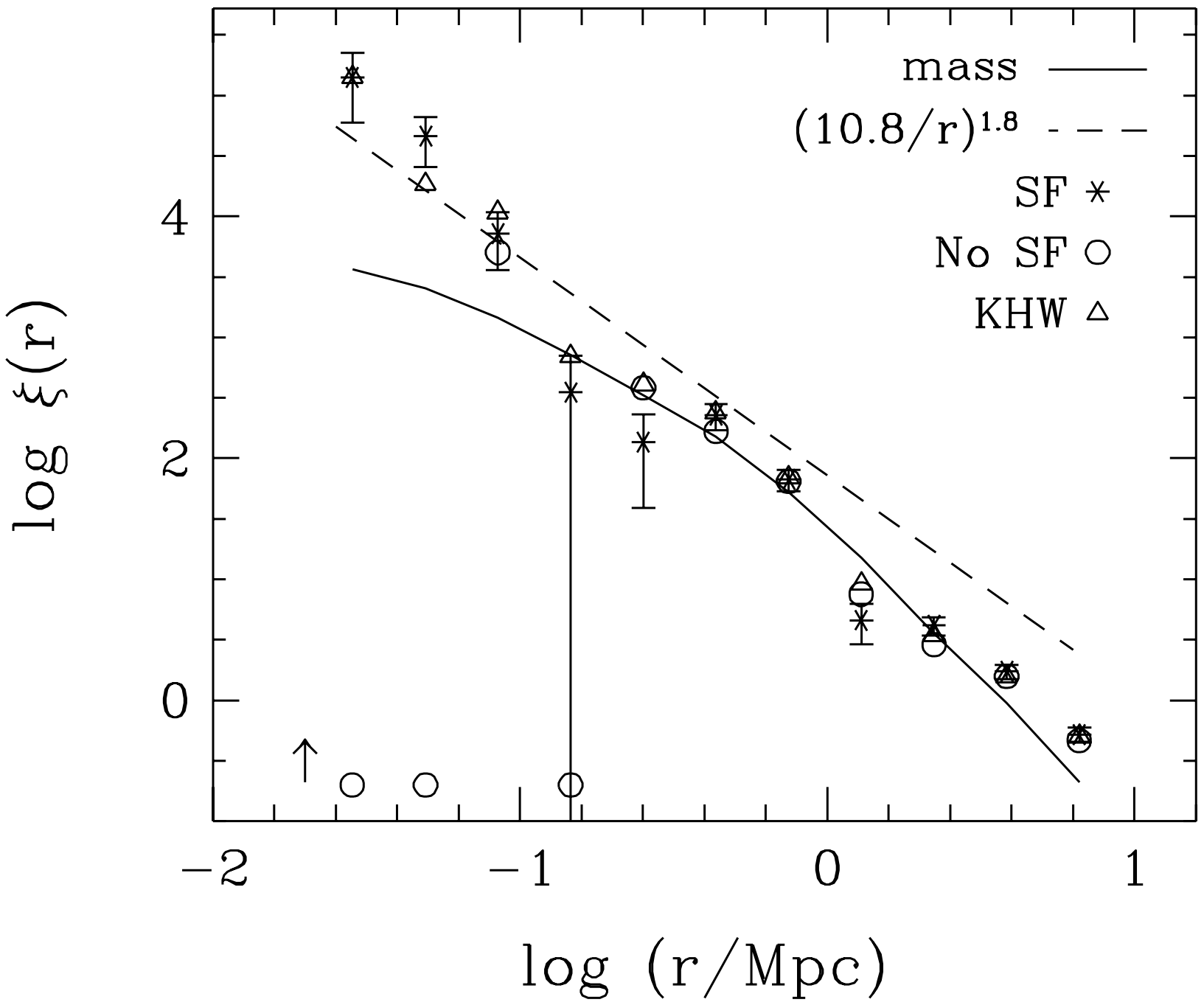

Fig. 7.— Correlation functions at $z = 0$. Asterisks show the correlation function of galaxies from the star-formation simulation of Figure 5b, with error bars representing the Poisson fluctuations in the number of galaxy pairs. Circles show the correlation function of galaxies from the simulation without star formation (Figure 5a); points near the x-axis indicate bins that contained no pairs. Triangles show the correlation function of galaxies from the KHW simulation. The solid line shows the mass correlation function from a simulation without gas. The dashed line shows the correlation function inferred by Davis & Peebles (1983) from the CfA redshift survey. The downward diagonal of the box is $(r/5.3\ \mathrm{Mpc})^{-2.1}$. The arrow in the lower-left corner indicates our gravitational softening scale of 20 kpc.

The simulations presented here confirm the most important qualitative conclusion of KHW, Katz & White (1993), and ESD: a realistic treatment of the baryon component yields dense, radiatively cooled objects that have roughly galactic masses and that survive as distinct entities after their dark matter halos merge. Quantitatively, we obtain similar results from these two simulations and from the instantaneous star-formation run of KHW. Because of the limited simulation volume, it is difficult to distinguish systematic differences between the simulations from small-number statistical fluctuations. At $z = 0.375$, the last hydrodynamic output of KHW's simulation, the new gas-only simulation is very similar, despite changes to the radiative cooling function, a more accurate treatment of the mean molecular weight in the equation of state (§3.2), and the imposition of a minimum SPH smoothing length (§2.3). This agreement is unsurprising, but it is a reassuring indication that simulation results are robust to small changes in the treatment of gas dynamics. We show elsewhere (Weinberg et al. 1995) that the galaxy population is also insensitive to assumptions about the photoionizing background, provided the simulation has enough mass resolution to track cooling in low-mass objects.

## 6. Concluding remarks

With simulations like those in §5, the relation between galaxies and mass becomes an *a priori* prediction of a theoretical model, instead of a source of additional freedom in the comparison to observed galaxy clustering. More and larger simulations will be required to reduce statistical uncertainties and to detail the dependence of the simulated galaxy population on parameters of the theoretical model, on numerical resolution of the simulation, and on assumptions about gas microphysics and star formation. Hydrodynamic simulations also follow the evolution of the intergalactic medium, so they can be tested against observations that range from X-ray emission by hot gas in galaxy groups and clusters to the absorption by cold gas in the spectra of background quasars. Quasar absorption data may play an especially important role over the next few years, in part because the Keck telescope and space-based UV instruments are rapidly improving the observational situation, and in part because these data probe high redshifts and small spatial scales, thereby complementing the information that comes from galaxy redshift surveys and microwave background studies.

Cosmological N-body simulations have played a variety of interlocking roles in the study of structure formation: offering insight into the physics of gravitational clustering, revealing the connections between initial conditions and evolved structure, guiding the development of statistical analysis methods, making quantitative predictions for specified theoretical models, and providing artificial data sets that aid the interpretation of observations. At a qualitative level, their most important achievement has been to show that small-amplitude fluctuations in a smooth, expanding universe can evolve into a network of clusters, filaments, sheets, and voids that bears a remarkable resemblance to observed large-scale structure. Quantitatively, they have begun to show how observational data constrain the answers to fundamental questions about the material contents of the universe, the values of cosmological parameters, and the origin of primordial inhomogeneities. Hydrodynamic simulations are a logical next step in these investigations because they include the most important of the physical processes that are neglected by the N-body approach. By so doing, they can close the gap between the simulated distribution of matter and the observable tracers of cosmic structure.

We have benefited from stimulating and informative discussions with numerous colleagues, especially Renyue Cen, Gus Evrard, Jeremiah Ostriker, Martin Rees, and Frank Summers. This work was supported in part by the Pittsburgh Supercomputing Center, the National Center for Supercomputing Applications (Illinois), the San Diego Supercomputing Center, the Alfred P. Sloan Foundation, a Hubble Fellowship, NASA Theory Grants NAGW-2422, NAGW-2523, and NAG5-2882, NASA HPCC/ESS Grant NAG 5-2213, and the NSF under Grants AST90-18526, ASC 93-18185 and the Presidential Faculty Fellows Program. DHW acknowledges the support of a Keck fellowship at the Institute for Advanced Study during much of this work.

---